\documentclass[aps,prd,
floatfix,nofootinbib,groupedaddress,showpacs]{revtex4-1}

\usepackage{amsmath}
\usepackage{amssymb}
\usepackage{graphicx}

\usepackage{latexsym}
\usepackage{array}
\usepackage{verbatim}

\usepackage{rotating}
\usepackage{color}

\usepackage[utf8]{inputenc}

\def\lsim{\mathrel{\raise.3ex\hbox{$<$\kern-.75em\lower1ex\hbox{$\sim$}}}}
\def\gsim{\mathrel{\raise.3ex\hbox{$>$\kern-.75em\lower1ex\hbox{$\sim$}}}}

\newcommand{\be}{\begin{equation}}
\newcommand{\ee}{\end{equation}}

\newlength{\absize}
\def\lsim{\mathrel{\rlap{\raise 2.5pt \hbox{$<$}}\lower 2.5pt
\hbox{$\sim$}}}

\newcommand{\TeV}{{\rm TeV}}
\newcommand{\Lumint}{{\cal L}_{\rm int}}

\newcommand{\sla}[1]{/\!\!\!\!#1}

\allowdisplaybreaks

\begin{document}

\vspace*{-1.5cm}
\begin{flushright}
DESY 18-170\\
\end{flushright}

\title{Improved constraints on the mixing and mass of $Z'$ bosons
from resonant diboson searches at the LHC at
$\sqrt{s}=13$ TeV and predictions for Run~II}

\author{I.~D. Bobovnikov}
\email{boboilya@yandex.by} \affiliation{
Deutsches Elektronen-Synchrotron DESY, Notkestrasse 85, Hamburg
22607, Germany}
\affiliation{The Abdus Salam ICTP Affiliated
Centre, Technical University of Gomel, 246746 Gomel, Belarus}
\author{P. Osland}
\email{Per.Osland@uib.no}
\affiliation{Department of Physics and
Technology, University of Bergen, Postboks 7803, N-5020 Bergen,
Norway}
\author{A.~A. Pankov}
\email{pankov@ictp.it}
\affiliation{The Abdus Salam ICTP
Affiliated Centre, Technical University of Gomel, 246746 Gomel,
Belarus}
\affiliation{Institute for Nuclear Problems, Belarusian
State University, 220030 Minsk, Belarus}
\affiliation{Joint
Institute  for Nuclear Research, Dubna 141980 Russia}

\date{\today}

\begin{abstract}

New neutral vector bosons $Z'$ decaying to charged gauge boson
pairs $W^+W^-$ are predicted in many scenarios of new physics,
including models with an extended gauge sector such as $E_6$,
left-right symmetric $Z^\prime_{\rm LRS}$ and the sequential
standard model $Z'_{\rm SSM}$. For these benchmark models we
calculate and present theoretical expectations for different
values of the $Z^\prime$ mass $M_2$ and mixing parameter $\xi$.
Our results are based on the narrow width approximation which
allows to make a convenient comparison of experiment to
theoretical benchmark models. The diboson production allows to
place stringent constraints on the $Z$-$Z'$ mixing angle and the
$Z'$ mass, which we determine by using data from $pp$ collisions
at $\sqrt{s}=13$ TeV recorded by the ATLAS detector at the
CERN LHC, with integrated luminosity of $\sim$ 36 fb$^{-1}$. By
comparing the experimental limits to the theoretical predictions
for the total cross section of $Z'$ resonant production and its
subsequent decay into $W^+W^-$ pairs, we show that the derived
constraints on the mixing angle for the benchmark models are of
the order of a few $\times 10^{-4}$, i.e., greatly improved with
respect to those derived from the global analysis of electroweak
data. We combine  the limits derived from diboson production
data with those obtained from the Drell--Yan process in order to
significantly extend the exclusion region in the $M_{2}$-$\xi$
parameter plane. Also, we demonstrate that further improvement on
the constraining of this mixing can be achieved through analysis
of the full set of Run~II data.

\end{abstract}

\maketitle

\section{Introduction}
\label{sec:I}

Neutral vector bosons, $Z^\prime$, are among the best motivated scenarios of
physics beyond the Standard Model (SM) \cite{Langacker:2008yv}.
Many new physics models beyond the SM \cite{Tanabashi:2018},
including superstring and left-right-symmetric models, predict the
existence of such bosons. They might actually be light enough
to be accessible at current and/or future colliders. The search
for such neutral $Z^{\prime}$  gauge bosons is an important
aspect of the experimental physics program of present and future
high-energy colliders.

Limits from direct production at the LHC and  virtual effects at
the Large Electron-Positron Collider (LEP), through interference
or mixing with the $Z$ boson, imply that any new $Z^{\prime}$
boson is rather heavy and mixes very little with the $Z$ boson.
Depending on the considered theoretical model, $Z^{\prime}$ masses
of the order of 4.5~TeV \cite{Aaboud:2017buh, Sirunyan:2018exx}
and $Z$-$Z^{\prime}$ mixing angles at the level of $10^{-3}$ are
already excluded~\cite{Erler:2009jh,delAguila:2010mx} (see
also \cite{Aaltonen:2010ws,Andreev:2012zza}). These constraints
come from the very high-precision $Z$ pole experiments at LEP and
the Stanford Linear Collider (SLC) \cite{ALEPH:2005ab}, including
measurements from the $Z$  line shape, from the leptonic branching
ratios (normalized to the total hadronic $Z$ decay width) as well
as from leptonic forward-backward asymmetries. While these
experiments were virtually blind to $Z'$ bosons with negligible
$Z$-$Z'$ mixing, precision measurements at lower and higher
energies (away from the $Z$ pole) attainable at TRISTAN
\cite{Pankov:1989bc} and LEP2 \cite{Osland:1996vv}, respectively,
were able to probe the $Z'$ exchange amplitude {\em via\/} its
interference with the photon and the SM $Z$ boson.

However, as was shown  in  \cite{Andreev:2014fwa}, the LHC at
nominal collider energy of $\sqrt{s}=14$ TeV and integrated
luminosity of $\Lumint=100$ fb$^{-1}$ has a high potential to
improve significantly on the current limits on the $Z$-$Z^\prime$ mixing
angle in the diboson channel
\begin{equation}\label{procWW}
pp \to (Z_2\to W^+ W^-)+X.
\end{equation}
This was demonstrated  in a recent paper
\cite{Osland:2017ema}
for the ``Altarelli Reference Model''
\cite{Altarelli:1989ff}, also known as the Sequential Standard
Model (SSM),
by using the current ATLAS
\cite{Aaboud:2017fgj} and CMS \cite{Sirunyan:2017acf} data
collected at a center of mass energy of $\sqrt{s}=13$ TeV
during searches for resonant $W^+W^-$ diboson production.
The SSM is often taken as a convenient benchmark by
experimentalists \cite{Benchekroun:2001je}. In this model, the new
heavy gauge bosons ${Z^\prime}_{\rm SSM}$ are considered heavy
carbon copies of the familiar $Z$, with the same coupling
constants.

In ATLAS $W^+W^-$ events are reconstructed via their semileptonic
decays, where one $W$ boson decays into a charged
lepton ($l=e,\mu$) and a neutrino, and the other into two jets
\cite{Aaboud:2017fgj}. CMS collects data where both $W$ bosons decay
hadronically with two reconstructed jets (dijet channel)
\cite{Sirunyan:2017acf}. The analysis presented below is based on
$pp$ collision data at a center-of-mass energy $\sqrt{s}=13$ TeV,
collected by the ATLAS experiment (36.1 fb$^{-1}$).
We consider here information provided by
ATLAS in published papers and in the HEPDATA
database \cite{hepdata}.
We shall also comment on the corresponding CMS result
\cite{Sirunyan:2017acf}.
The data is used to probe
the $Z$-$Z^{\prime}$ mixing and to interpret the results of the present
analysis within models with an extended gauge sector. Among these,
models based on the $E_6$ GUT group and left-right symmetry groups
have been extensively pursued in the literature and are
particularly significant from the point of view of LHC
phenomenology. Here we extend our analysis presented in
\cite{Osland:2017ema} for the SSM to various $Z'$ models, which
include the $E_6$ based $Z^{\prime}_\chi$, $Z^{\prime}_\psi$,
$Z^{\prime}_\eta$, and also the $Z^{\prime}_{\rm LRS}$ boson
appearing in models with left-right symmetry. Thus, our present
analysis is complementary to the previous studies
\cite{Osland:2017ema}.

One should emphasize that we made a choice of
particular benchmark models to represent different qualitative
features of those $Z'$ models, such as the fact that those models
typically involve an extra neutral $Z'$ boson with relatively narrow
width (which however may become larger if non-SM particles are
included in the decays in addition to the SM states).

The $W^+W^-$ pair production process (\ref{procWW}) is
very important for diagnostics of the electroweak gauge symmetry.
General properties of the weak gauge bosons are closely related to
electroweak symmetry breaking and to the structure of the gauge
sector, like the existence and structure of trilinear couplings.
Also, the diboson decay mode of the $Z'$ probe the gauge
coupling strength between the new and the SM gauge bosons
\cite{Pankov:1992cy,Andreev:2012cj,Andreev:2014fwa,Andreev:2015nfa,Osland:2017ema}.
In addition, the coupling strength strongly influences the decay
branching ratios and the natural widths of such a new gauge boson.
Thus, detailed examination of the process (\ref{procWW}) will not only
test the gauge sector of the SM with high accuracy, it will also shed light
on any new physics (NP) that may appear beyond the SM. Here, we
examine the feasibility of observing $Z^{\prime}$ boson effects
in the $W^+W^-$ pair production process at the LHC.

In contrast to the Drell-Yan (DY)  process
\begin{equation}
pp \to Z' \to \ell^+ \ell^-+X, \label{procDY}
\end{equation}
with $\ell=e, \mu$, the diboson process is not the principal discovery channel, but
can help to understand the origin of new gauge bosons.

At Tevatron energies, direct searches for heavy $W^+W^-$~resonances have been performed
by both the CDF and D0 collaborations. The D0
collaboration explored diboson resonant production up to masses $\sim{\cal
O}(700~\text{GeV})$ using the pure leptonic $\ell\nu \ell' \nu'$
and semi-leptonic $\ell \nu j j$ final states
\cite{Abazov:2010dj}. On the other hand, the CDF collaboration searched for
resonant $W^+W^-$ production in the $e \nu j j$ final state, resulting
in a lower limit on the masses of $Z'$ and $W'$
bosons~\cite{Aaltonen:2010ws}, excluding masses up to ${\cal
O}(900~\text{GeV})$, depending on the mixing.

Previous searches for diboson ($VV$) resonances at the LHC were carried
out by the ATLAS and CMS collaborations with $pp$ collisions at
$\sqrt{s} = 7$, 8 and 13~\TeV. These include fully leptonic
($\ell\nu \ell\nu$, $\ell\nu
\ell\ell$)~\cite{Aad:2012nev,Collaboration:2012iua,Aad:2014pha,Khachatryan:2014xja},
semileptonic ($\nu\nu jj$, $\ell\nu jj$, $\ell\ell jj$)
\cite{Aaboud:2016okv,Khachatryan:2014gha,Sirunyan:2016cao}
and fully hadronic ($jjjj$) $VV$ \cite{Aaboud:2016okv,Sirunyan:2016cao}
final states. By
combining the results of searches in the $\nu\nu jj$, $\ell\nu
jj$, $\ell\ell jj$ and $jjjj$ channels, the ATLAS
Collaboration~\cite{Aaboud:2016okv} set a lower bound of
2.60~\TeV{} on the mass of a spin-1 resonance at the 95\%
confidence level, in the context of the heavy vector triplet model.
The recent results presented in
\cite{Aaboud:2017fgj,Sirunyan:2017acf} by the ATLAS and CMS
collaborations using, respectively, semileptonic  and hadronic
final-state events in $pp$ collision data at 13~TeV benefit from
an integrated luminosity of $\sim$ 36 fb$^{-1}$, which is an order
of magnitude larger than what was available for the previous search in
the fully hadronic final state at $\sqrt{s} =
13~\TeV$~\cite{Aaboud:2016okv}.

It should be noted that the future $e^+e^-$ International linear
collider (ILC) with high c.m.\ energies and longitudinally
polarized beams could indicate the existence of $Z^{\prime}$
bosons via its interference effects in fermion pair production
processes, with masses up to about $6\times \sqrt{s}$
\cite{Rizzo:2006nw} while $Z$-$Z'$ mixing will be constrained down
to $\sim 10^{-4}-10^{-3}$ in the process $e^+e^-\to W^+W^-$
\cite{Andreev:2012cj}.

In this work, we derive bounds on a possible new neutral spin-1
resonance ($Z^\prime$) for the considered models  from the available
ATLAS data on $W^+W^-$ pair production
\cite{Aaboud:2017fgj}. We present results as
constraints on the relevant $Z$-$Z^{\prime}$ mixing angle
introduced in Sect.~II and on the $M_{Z^\prime}$ mass.

The paper is organized as follows. In
Section~\ref{sect-ZZ'mixing}, we briefly describe the $Z$-$Z'$
mixing and emphasize its role in the process (\ref{procWW}). In
Sec.~\ref{sect-cross} we summarize the relevant cross section, in
Sec.~\ref{sec:width} we study the $Z_2\to W^+W^-$ width and then
in Sec.~\ref{sect-constraints} we show the resulting constraints
on the $M_2$-$\xi$ parameter space, whereas in Sec.~\ref{sect-DY}
we discuss results from the DY process $q\bar q\to Z_2\to
l^+l^-$. In Sec.~\ref{sect:overall_constraints} we collect and
compare constraints from the diboson process with those deduced
from the Drell--Yan process, and in Sec.~\ref{sect:conclusions} we
conclude.

\section{$Z$-$Z'$ mixing}
\label{sect-ZZ'mixing}

Signals of $Z'$ in representative models such as the $E_6$
models, the LR model and the SSM have been extensively studied in the literature
\cite{Langacker:2008yv} and explored at LEP2, the Tevatron and the
LHC. For the present notation we refer to \cite{Andreev:2012cj}, where
also a brief description of the models can be found.

The mass-squared matrix of the $Z$ and $Z^{\prime}$ can have
non-diagonal entries $\delta M^2$, which are related to the vacuum
expectation values of the fields of an extended Higgs sector:
\begin{equation}\label{massmatrix}
M_{ZZ^\prime}^2 = \left(\begin{matrix} M_Z^2 & \delta M^2\\ \delta
M^2 & M_{Z^\prime}^2
\end{matrix}\right).
\end{equation}
Here, $Z$ and $Z^{\prime}$ denote the weak gauge boson eigenstates
of $SU(2)_L\times U(1)_Y$ and of the extra $U(1)'$, respectively.
The mass eigenstates, $Z_1$ and $Z_2$, which diagonalize the matrix
(\ref{massmatrix}), are obtained by a rotation of the
fields $Z$ and $Z^\prime$:
\begin{subequations}
\label{Eq:Z12-couplings}
\begin{eqnarray}
&& Z_1 = Z\cos\phi + Z^\prime\sin\phi\;, \label{z1} \\
&& Z_2 = -Z\sin\phi + Z^\prime\cos\phi\;. \label{z2}
\end{eqnarray}
\end{subequations}
The mixing angle $\phi$ is expressed in terms of masses as \cite{Langacker:2008yv}:
\begin{equation}
\label{phi} \tan^2\phi={\frac{M_Z^2-M_1^2}{M_2^2-M_Z^2}}\simeq
\frac{2 M_Z \Delta M}{M_2^2}\;,
\end{equation}
where the downward shift $\Delta M=M_Z-M_1>0$, $M_Z$ being the
mass of the $Z_1$ boson in the absence of mixing, i.e., for
$\phi=0$. We assume the mass $M_1$ is determined
experimentally, the mixing then depends on two free parameters, which
we identify as $\phi$ and $M_2$, and we will adopt this
parametrization throughout the paper. In the important
limit  $M_{Z'} \gg (M_{Z}, \Delta M)$ one finds
\cite{Langacker:2008yv}
\begin{equation}
M_1 \sim M_{Z}  \ll M_{Z'} \sim M_2. \label{MZ}
 \end{equation}

The mixing angle $\phi$ will play an important role in our
analysis. Such mixing effects reflect the underlying gauge
symmetry and/or the Higgs sector of the model. For instance, in
certain models one specifies not only the $SU(2)$ assignments but the
$U(1)'$ assignments of the Higgs fields. To a good approximation,
for $M_1\ll M_2$, in specific ``minimal Higgs'' models, one has an
additional constraint \cite{Langacker:1991pg}
\begin{equation}\label{phi0}
\phi\simeq -s^2_\mathrm{W}\
\frac{\sum_{i}\langle\Phi_i\rangle{}^2I^i_{3L}Q^{\prime}_i}
{\sum_{i}\langle\Phi_i\rangle^2(I^i_{3L})^2} ={\cal C}\
{\frac{\displaystyle M^2_1}{\displaystyle M^2_2}},
\end{equation}
where $s_\mathrm{W}$ is the sine of the electroweak  angle.
In these models $\phi$ and $M_2$ are not independent and
there is only one (e.g., $M_2$) free parameter. Furthermore,
$\langle\Phi_i\rangle$ are the Higgs vacuum expectation values
spontaneously breaking the symmetry, and $Q^\prime_i$  are their
charges with respect to the additional $U(1)'$. In
these models the same Higgs multiplets are responsible for both
generation of the mass $M_1$ and for the strength of the
$Z$-$Z^\prime$ mixing. Thus ${\cal C}$ is a model-dependent
constant.

From (\ref{Eq:Z12-couplings}), one obtains the vector and
axial-vector couplings of the $Z_1$ and $Z_2$ bosons to fermions:
\begin{subequations}
\begin{eqnarray}
&&\hspace{-10mm} v_{1f} = v_f\cos\phi + v_f^\prime
\sin\phi\;,\;a_{1f} = a_f \cos\phi +
a_f^\prime \sin\phi\;, \label{v1} \\
&&\hspace{-10mm} v_{2f} = v_f^\prime \cos\phi - v_f
\sin\phi\;,\;a_{2f} = a_f^\prime \cos\phi -a_f \sin\phi\;,
\label{v2}
\end{eqnarray}
\end{subequations}
with $v_f^\prime$ and $a_f^\prime$ the fermionic $Z^\prime$
couplings which can be found, e.g., in~\cite{Andreev:2012cj}.

We will consider NP models where $Z^\prime$'s interact with
charged gauge bosons $W^\pm$ via their mixing with the SM $Z$,
assuming that the $Z^\prime$ couplings exhibit the same Lorentz
structure as those of the SM. An important property of the models
under consideration is that the gauge eigenstate $Z^{\prime}$ does
not couple to the $W^+W^-$ pair since it is neutral under
$SU(2)_L$. Therefore the process (\ref{procWW}) is sensitive to a
$Z^\prime$ only in the case of a non-zero $Z$-$Z^{\prime}$ mixing.
From (\ref{z1}) and (\ref{z2}), one obtains:
\begin{subequations}
\begin{eqnarray}
&& g_{WWZ_1}=\cos\phi\;g_{WWZ}\;, \label{WWZ1} \\
&& g_{WWZ_2}=-\sin\phi\; g_{WWZ}\;,\label{WWZ2}
\end{eqnarray}
\end{subequations}
where $g_{WWZ}=\cot\theta_W$\footnote{In our analysis, we ignore
kinetic mixing \cite{Holdom:1985ag}. Such mixing would introduce
an additional parameter, and could modify the exclusion reach
(see, for example \cite{Krauss:2012ku,Hirsch:2012kv}).}. Also,
$g_{WW\gamma}=1$.

In many extended gauge models, while the couplings to fermions are
not much different from those of the SM, the $Z_2WW$ coupling is
substantially suppressed with respect to that of the SM. In fact,
in the extended gauge models  the SM trilinear gauge boson
coupling strength, $g_{WWZ}$, is replaced by $g_{WWZ} \rightarrow
\xi\cdot g_{WWZ}$, where $\xi \equiv \vert\sin\phi\vert$
(see Eq.~(\ref{WWZ2})) is the mixing factor. We will set cross section
limits on such $Z_2$ as functions of the mass $M_2$ and $\xi$.

\section{Cross section }
\label{sect-cross}

The differential cross section for  the process (\ref{procWW})
from initial quark-antiquark states can be written as
\begin{eqnarray}
 \frac{d\sigma}{dM\,dy\,dz}
 = K \frac{2 M}{s}
\sum_q [f_{q|P_1}(\xi_1)f_{\bar q|P_2}(\xi_2) + f_{\bar
q|P_1}(\xi_1)f_{q|P_2}(\xi_2)]\, \frac{d\hat \sigma_{q \bar
q}}{dz}. \label{dsigma}
\end{eqnarray}
Here, $s$ denotes the proton-proton center-of-mass energy squared,
$z\equiv\cos\theta$, with $\theta$ the $W^-$-boson--quark angle in
the $W^+W^-$ center-of-mass frame, $y$ the diboson rapidity and
$M$ the diboson $W^+W^-$ invariant mass. Furthermore,
$f_{q|P_1}(\xi_{1},M)$ and $f_{\bar q|P_2}(\xi_{2},M)$ are quark
and antiquark distribution functions for the protons $P_1$ and
$P_2$, respectively, with $\xi_{1,2}=(M/\sqrt s)\exp(\pm y)$ the
parton fractional momenta. Finally, $d\hat \sigma_{q \bar q}/dz$
are the partonic differential cross sections, to be specified
below. In~(\ref{dsigma}), the $K$ factor accounts for higher-order
QCD contributions.
 For numerical computation, we use the CTEQ-6L1 parton distributions
\cite{Pumplin:2002vw}. Our estimates will be at the Born level,
thus the factorisation scale $\mu_{\rm F}$ enters only through the
parton distribution functions, as the parton-level cross section
at this order does not depend on $\mu_{\rm F}$. As regards the
scale dependence of the parton distributions we choose for the
factorization scale the $W^+W^-$ invariant mass, $\mu_{\rm
F}^2=M^2=\hat{s}$, with $\hat{s}=\xi_1 \,\xi_2\,s$  the parton
subprocess c.m.\ energy squared. The obtained constraints
presented in the following are numerically not significantly
modified when $\mu_{\rm F}$ is varied in the range from $\mu_{\rm
F}/2$ to $2\mu_{\rm F}.$

\begin{figure}[htb]
\begin{center}
\includegraphics[scale=0.75]{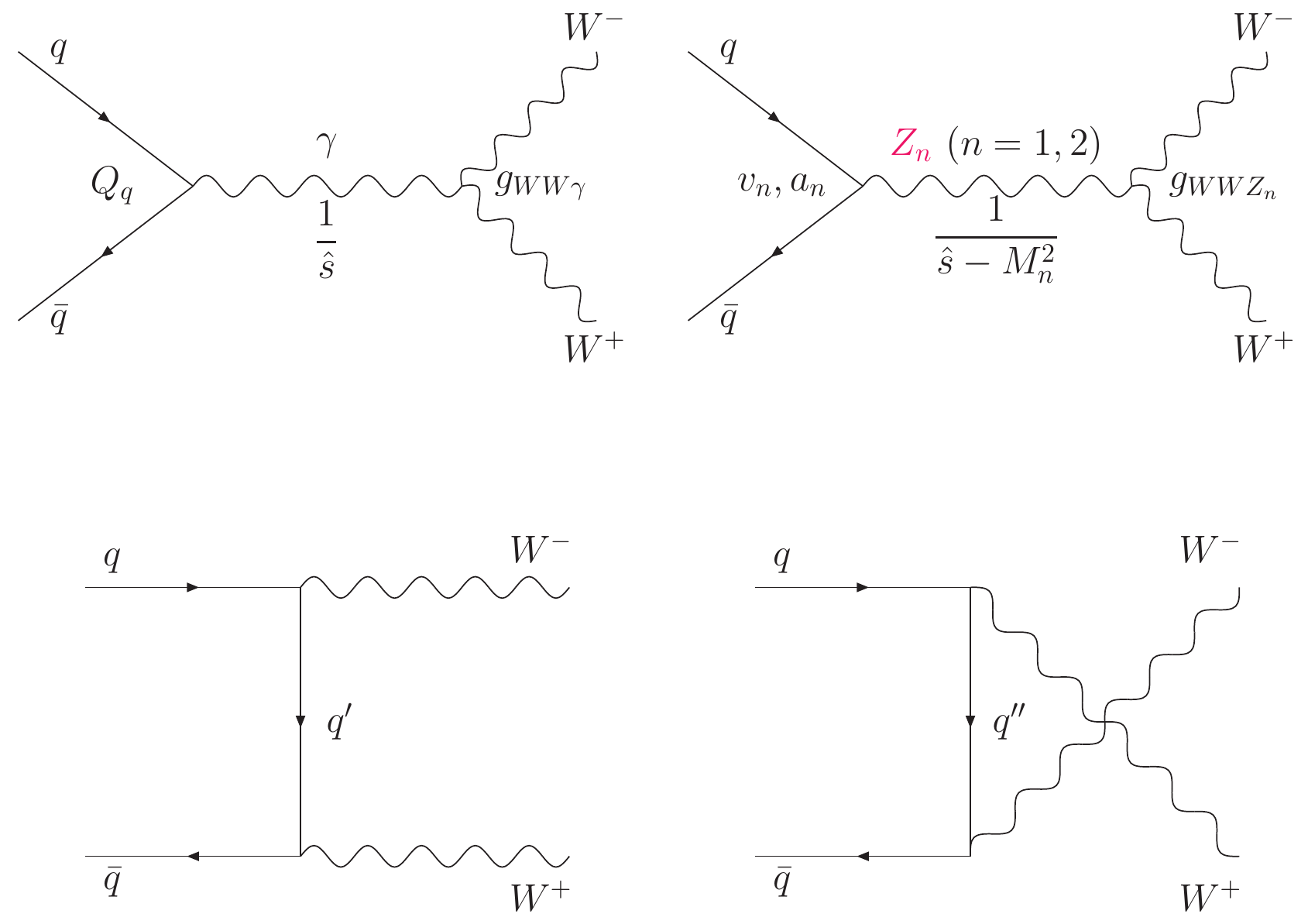}
\end{center}
\caption{Leading-order Feynman diagrams for the $q\bar{q} \to
{\gamma,Z_1,Z_2} \to  W^+W^-$ process  within the framework of
extended gauge models.}\label{fig1}
\end{figure}

The parton-level $W^+W^-$ boson pair production can be described,
within the gauge models discussed here, by the subprocesses
\cite{Andreev:2014fwa}
\begin{equation}
q\bar{q} \to {\gamma,Z_1,Z_2} \to W^+W^-,\label{parton}
\end{equation}
as well as $t$- and $u$-channel Feynman diagrams displayed in
Fig.~\ref{fig1}.

The differential cross section for the processes $q\bar{q}\to
W^+W^-$ described by Feynman diagrams depicted in Fig.~\ref{fig1}
and averaged over quark colors, can be written as
\cite{Tanabashi:2018}
\begin{eqnarray}
&&\left(\frac{\pi{\alpha_{\rm em}^2} \beta_W}{{N_C}\,
\hat{s}}\right)^{-1} \frac{d{\hat\sigma_{{q}\bar{q}}}}{dz} \nonumber \\
&=&[(Q_q+v_{1q}g_{WWZ_1}\chi_1+v_{2q}g_{WWZ_2}\chi_2)^2
+(a_{1q}g_{WWZ_1}\chi_1+a_{2q}g_{WWZ_2}\chi_2)^2]A(\hat s,\hat t,\hat u)\nonumber \\
&+& \frac{1}{2s_W^2}[Q_q+(v_{1q}+a_{1q})g_{WWZ_1}\chi_1 +
(v_{2q}+a_{2q})g_{WWZ_2}\chi_2]
[\theta(-Q_q)I(\hat s,\hat t,\hat u)-\theta(Q_q)I(\hat s,\hat u,\hat t)]\nonumber \\
&+&\frac{1}{8s_W^4}[\theta(-Q_q)E(\hat s,\hat t,\hat u)
+\theta(Q_q)E(\hat s,\hat u,\hat t)], \label{parton_cross}
\end{eqnarray}
where $\alpha_{\rm em} = 1/128.9$,  $\theta(x)=1$ for $x>0$ and
$\theta(x)=0$ for $x<0$, $N_C$ being the color factor  ($N_C=3$
for quarks), and
\begin{eqnarray}
&&A(\hat s,\hat t,\hat
u)=\left(\frac{\hat{t}\hat{u}}{M_W^4}-1\right)\left(\frac{1}{4}-\frac{M_W^2}{\hat{s}}+
3\frac{M_W^4}{\hat{s}^2}\right)+\frac{\hat{s}}{M_W^2}-4,\nonumber \\
&&I(\hat s,\hat t,\hat
u)=\left(\frac{\hat{t}\hat{u}}{M_W^4}-1\right)\left(\frac{1}{4}-\frac{M_W^2}{2\hat{s}}-
\frac{M_W^4}{\hat{s}\hat{t}}\right)+\frac{\hat{s}}{M_W^2}-2+2\frac{M_W^2}{\hat{t}},\nonumber \\
&&E(\hat s,\hat t,\hat
u)=\left(\frac{\hat{t}\hat{u}}{M_W^4}-1\right)\left(\frac{1}{4}+\frac{M_W^4}{\hat{t}^2}\right)+\frac{\hat{s}}{M_W^2}.
\label{kinemat}
\end{eqnarray}
Here,  $\hat{s}$, $\hat{t}$, $\hat{u}$ are the Mandelstam
variables defined as $\hat{s}=M^2$,  $\hat{t}=M_W^2-\hat{s}(1
-\beta_W z)/2$, $\hat{u} =M_W^2-\hat{s}(1 + \beta_W z)/2 $;
$\chi_1={\hat{s}}/({\hat{s}-M_1^2+i M_1\Gamma_1})$,
$\chi_2={\hat{s}}/({\hat{s}-M_2^2+i M_2\Gamma_2})$, $\Gamma_{1,2}$
are total $Z_{1,2}$ boson decay widths; and $\gamma_{W} =
\sqrt{\hat{s}}/2M_W$. In the $t$- and $u$-channel exchanges of
Fig.~\ref{fig1} we account for the initial $q = u,d,s,c$, only the
CKM favoured quarks in the approximation of unity relevant matrix
element. The differential cross section for the processes
$q\bar{q}\to  W^+W^-$ in the SM  \cite{Tanabashi:2018} can be
reproduced from Eq.~(\ref{parton_cross}) if one ignores the
effects of the $Z$-$Z'$ mixing.

The differential cross section for the process $q\bar{q}\to Z_2\to
W^+W^-$, averaged over quark colors, can now be obtained from
Eq.~(\ref{parton_cross}) and written as \cite{Andreev:2014fwa}
\begin{align}
 \frac{d\hat{\sigma}^{Z_2}_{q \bar q}}{d \cos\theta}
 &= \frac{1}{3}\,\frac{\pi\alpha_{\rm em}^2 \cot^2\theta_W}{16 \,\hat{s}}
\beta_W^3\left(v_{2,f}^2 + a_{2,f}^2\right) \,
\vert\chi_2\vert^2 \nonumber \\
&  \times    \left(\frac{\hat{s}^2}{M_W^4} \sin^2\theta +
4\frac{\hat{s}}{M_W^2}(4-\sin^2\theta)+12\sin^2\theta\right)\cdot\xi^2.
\label{xsection2}
\end{align}
The resonant production cross section of process (\ref{procWW}) at
the hadronic level can be derived from Eqs.~(\ref{dsigma})
 and (\ref{xsection2}). Specifically, the total cross section for the narrow $Z_2$ state
is derived from (\ref{dsigma}) by integrating the right-hand side
over  the full phase space. In the narrow width approximation
(NWA),  one obtains \cite{Andreev:2014fwa}:
\begin{equation}
\sigma^{Z_2}  = \sigma(pp\to Z_2) \times \text{Br}(Z_2 \to W^+W^-)
\;, \label{TotCr}
\end{equation}
where $\sigma(pp\to Z_2) \times \text{Br}(Z_2 \to W^+W^-)$ is the
(theoretical) total $Z_2$ production cross section times branching
ratio determined in the total phase space.

However, it turns out that the real situation is more complicated
because the cross section for $W^+W^-$ pair production is measured
indirectly via decay products of $W$'s. In fact, the analysis
performed in Sect.~\ref{sect-constraints} is based on the
available ATLAS data of $W^+W^-$ pair production with their
subsequent decay into semileptonic final states
\cite{Aaboud:2017fgj} where one $W$ boson decays leptonically ($W
\to \ell \nu$ with $\ell = e, \mu$) and the other $W$ boson decays
hadronically ($W \to q\bar{q}^\prime$ with $q,q^\prime$ quarks)
which can be written as  $p p \to Z_2 \to W^+ W^- \to \ell^\pm jj
\, \sla{E}_T $ where $j$  stands for jets.

The cross section  is measured in a fiducial phase space and also
in the total phase space (see, e.g.
\cite{Aad:2012nev,ATLAS:2012mec}). The fiducial cross section
$\sigma_{\rm fid}^{Z_2}$ for the $p p \to Z_2 \to W^+ W^- \to
\ell^\pm jj \, \sla{E}_T $ process is calculated according to the
equation
\begin{equation}
\sigma_{\rm fid}^{Z_2}= \frac{N_{\rm data} - N_{\rm
bkg}}{\varepsilon \times {\Lumint}}= \frac{N^{Z_2}}{\varepsilon
\times {\Lumint}} \ , \label{fid}
\end{equation}
where $N_{\rm data}$ and $N_{\rm bkg}$ are the number of observed
data events and estimated background events, respectively,
$N^{Z_2}$ the number of signal events for a narrow $Z_2$ resonance
state, $\varepsilon$ is defined as the ratio of the number of
events satisfying all selection criteria to the number of events
produced in the fiducial phase space and is estimated from
simulation. $\Lumint$ is the integrated luminosity of the data
sample.

The total cross section $\sigma^{Z_2}$ for the $pp\rightarrow Z_2
\rightarrow W^+W^- + X$ process is calculated for each channel
 using the equation
\begin{equation}
 \sigma^{Z_2} = \frac{N^{Z_2}}{ \varepsilon\times
 A\times{\rm BR}\times {\Lumint}}\ , \label{tot}
\end{equation}
where $A$ represents the kinematic and geometric acceptance from
the total phase space to the fiducial phase space, and ${\rm BR}$
is the branching ratio for both $W$ bosons decaying into $l\nu
\bigoplus jj$. In other words, the overall acceptance times
trigger, reconstruction and selection efficiencies
($A\times\varepsilon$) is defined as the number of signal events
passing the full event selection divided by the number of
generated events. The total cross section $\sigma^{Z_2}$ is this
physical quantity that is measured experimentally at the LHC and
which will be used in our analysis performed in
Sect.~\ref{sect-constraints}.

\section{The $Z_2$ width}
\label{sec:width}
In the calculation of the total width $\Gamma_{2}$ we consider the
following channels: $Z_2\to f\bar f$, $W^+W^-$, and $Z_1H$
\cite{Barger:2009xg}, where $H$ is the SM Higgs boson and $f$ are
the SM fermions ($f=l,\nu,q$). Throughout the paper we shall
ignore the couplings of the $Z_2$ to beyond-SM particles such as
right-handed neutrinos, SUSY partners and any exotic fermions in
the theory, which all together may increase the width of the $Z_2$
by up to about a factor of five \cite{Kang:2004bz} and hence lower
the branching ratio into a $W^+W^-$ pair by the same factor.

\begin{table}[htb]
\caption{Ratio $\Gamma_{2}^{ff}/M_{2}$ for the $\chi, \psi, \eta$,
${\rm LRS}$ and SSM models.}
\begin{center}
\begin{tabular}{|c|c|}
\hline $Z_2$  & $\Gamma_{2}/M_{2}$ [\%] \\
\hline $\chi$ & 1.2 \\
\hline $\psi$ & 0.5 \\
\hline $\eta$ & 0.6 \\
\hline ${\rm LRS}$ & 2.0 \\
\hline SSM & 3.0 \\
\hline
\end{tabular}
\end{center}
\label{tab1}
\end{table}

\begin{figure}[htb]
\refstepcounter{figure} \label{fig2} \addtocounter{figure}{-1}
\begin{center}
\includegraphics[scale=0.30]{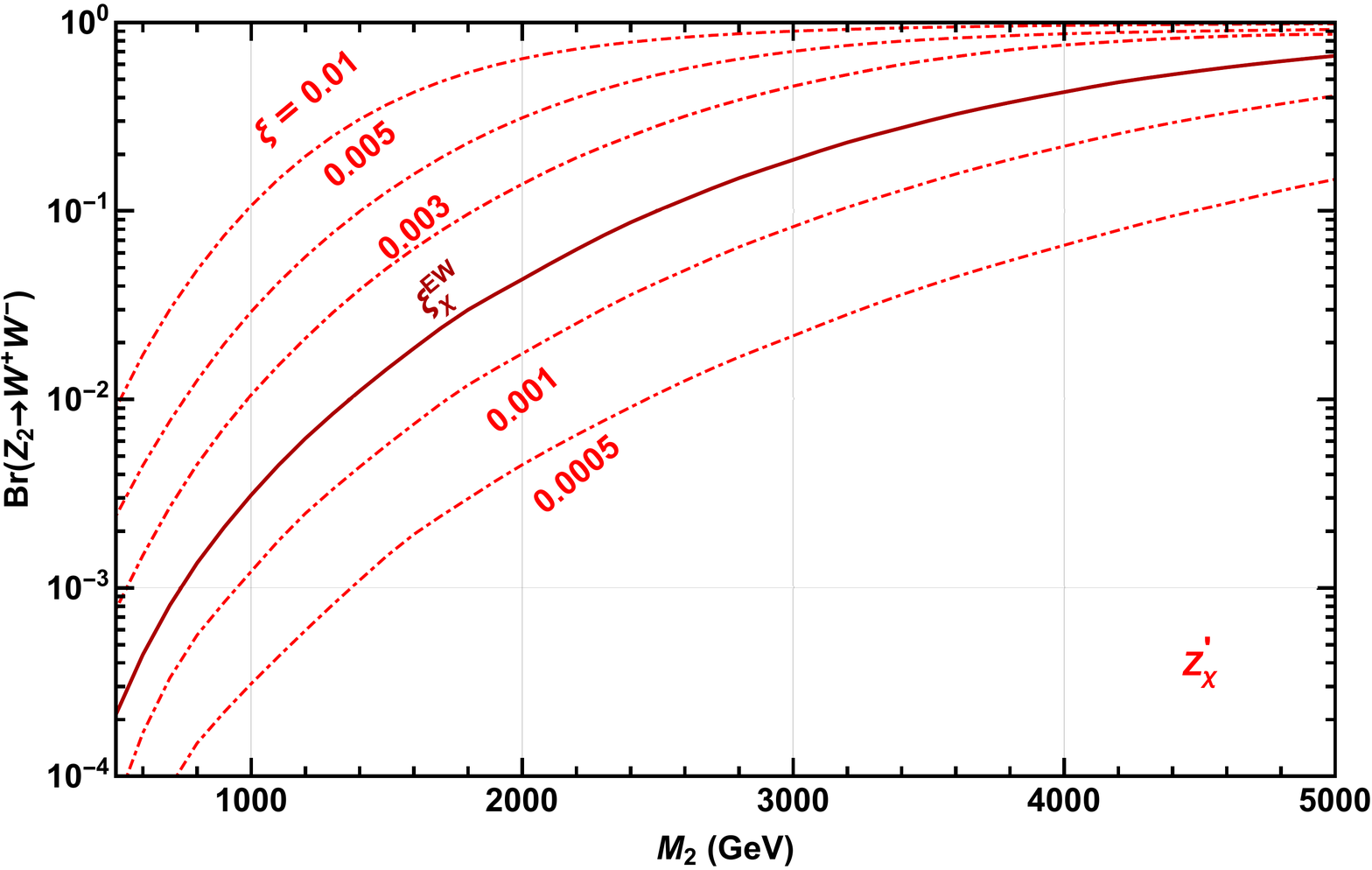}
\includegraphics[scale=0.30]{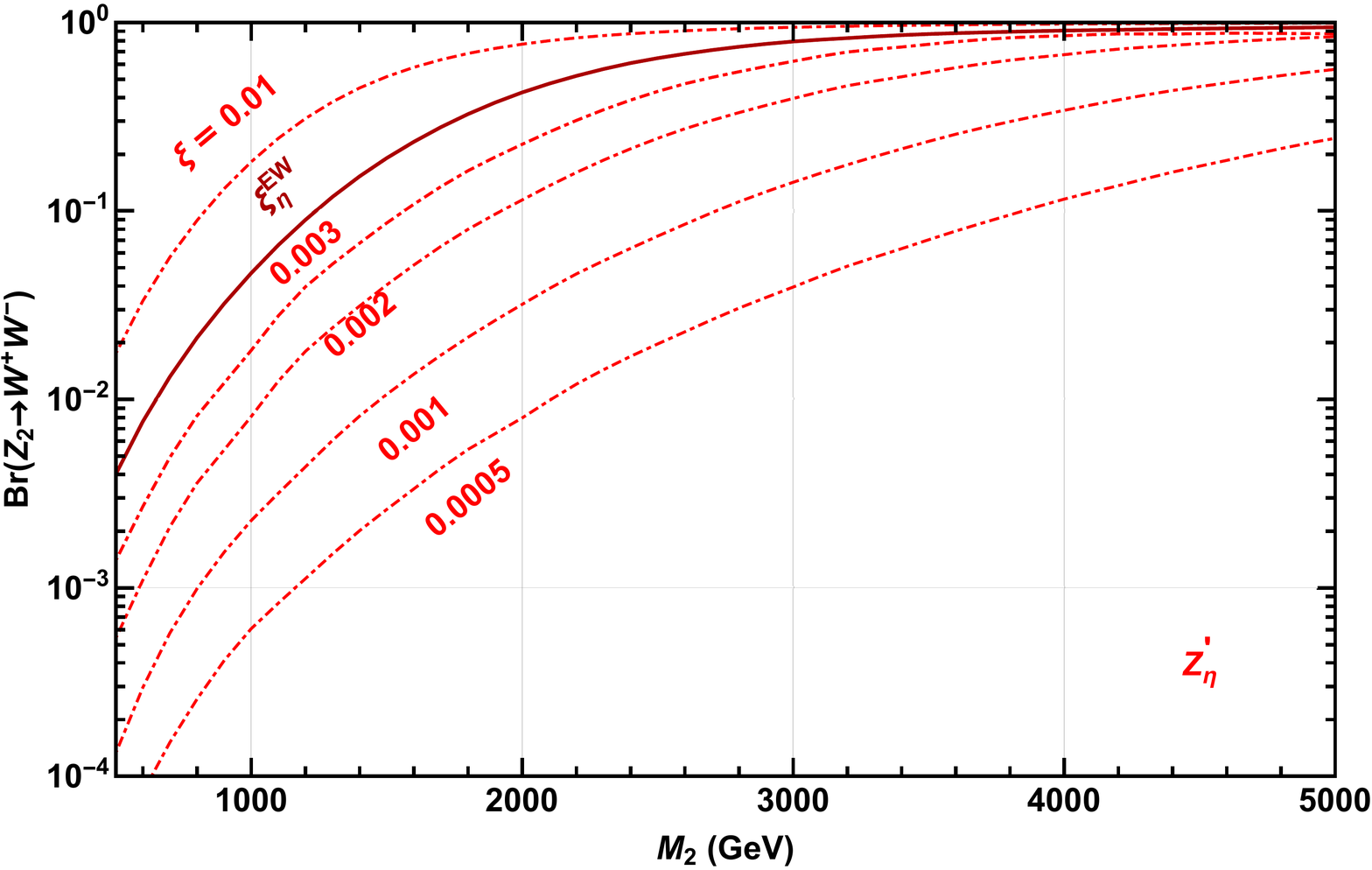}
\end{center}
\vspace*{-5mm}
\caption{ Branching fraction $\text{Br}(Z_2\to W^+W^-)$ vs
$M_2$ for the $\chi$ model (left panel) and the $\eta$ model (right
panel). Labels attached to the  curves  correspond to  a range of
values of  the mixing factor $\xi$ from 0.01 and down to 0.0005,
where values larger than the limits
obtained from the electroweak precision data, $\xi_\chi^{\rm
EW}=1.6\cdot 10^{-3}$ in the left panel and $\xi_\eta^{\rm
EW}=4.7\cdot 10^{-3}$ in the right panel, are shown only for
illustrative purposes. }
\end{figure}

\begin{figure}[htb]
\refstepcounter{figure} \label{fig3} \addtocounter{figure}{-1}
\begin{center}
\includegraphics[scale=0.30]{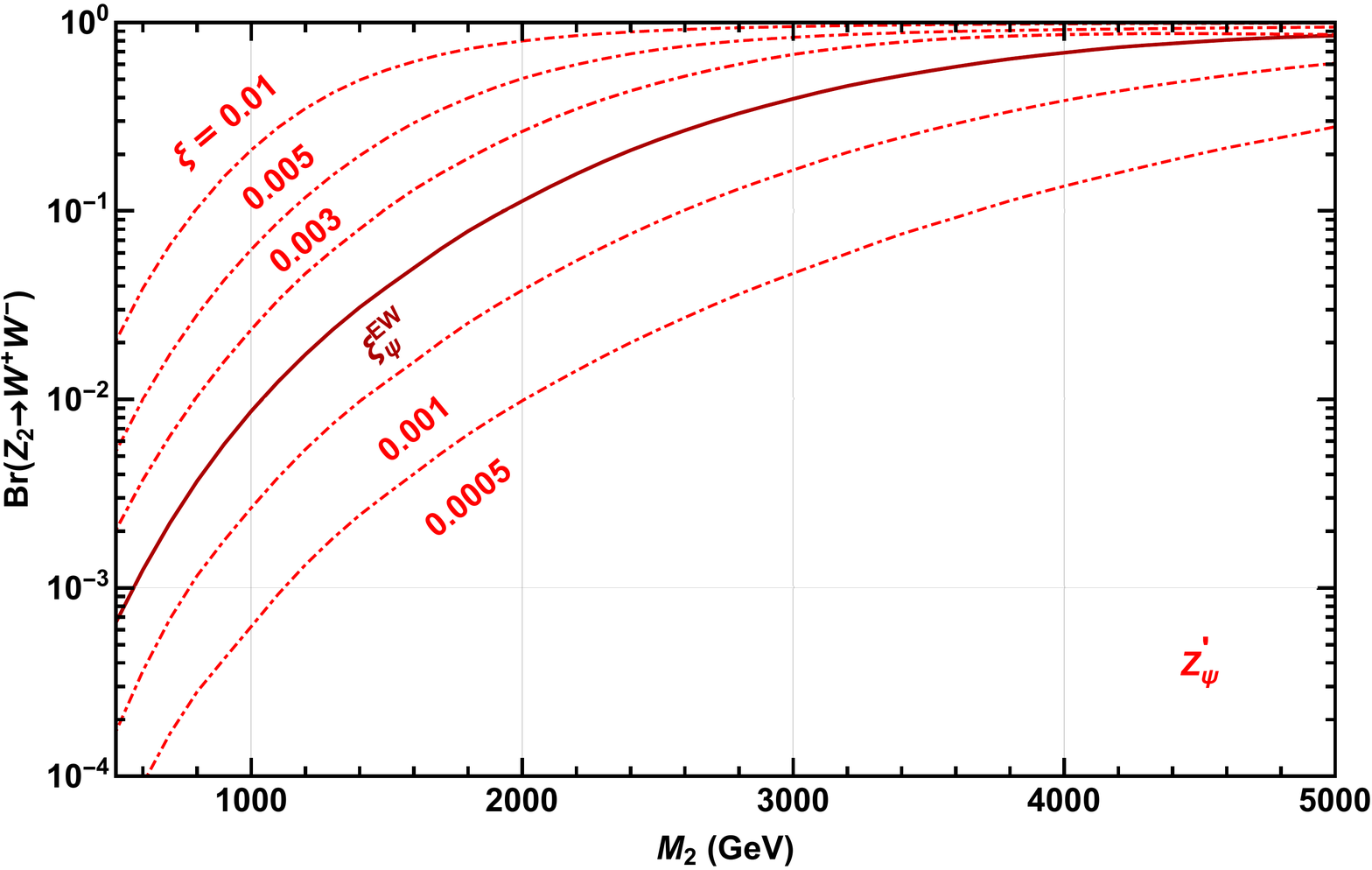}
\includegraphics[scale=0.30]{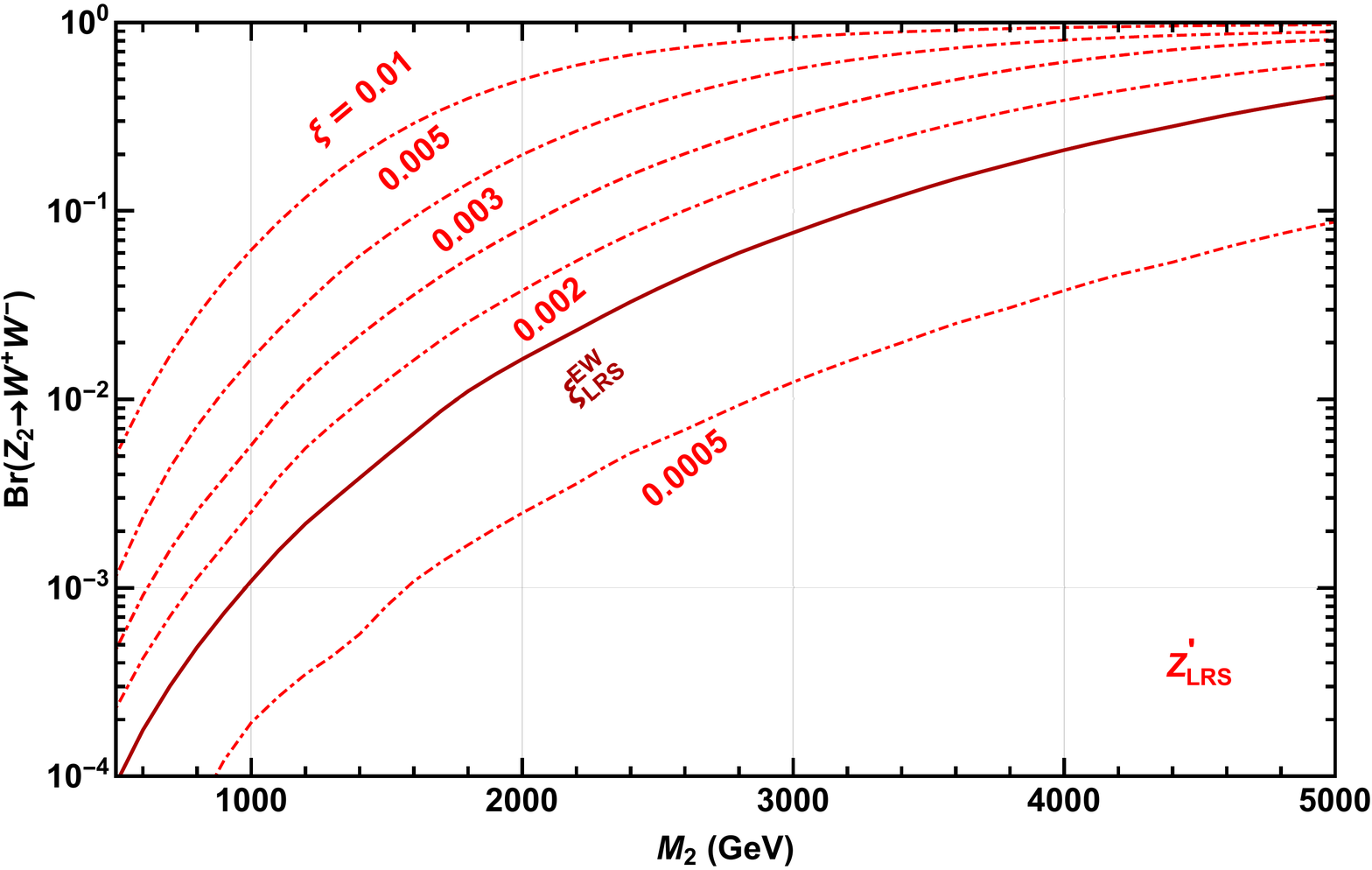}
\end{center}
\vspace*{-5mm}
\caption{Same as in Fig.~\ref{fig2} but for the $\psi$ model with
$\xi_\psi^{\rm EW}=1.8\cdot 10^{-3}$ (left panel) and for the ${\rm
LRS}$ model with $\xi_{\rm LRS}^{\rm EW}=1.3\cdot 10^{-3}$ (right
panel).}
\end{figure}

\begin{figure}[htb]
\refstepcounter{figure} \label{fig4}
\addtocounter{figure}{-1}
\begin{center}
\includegraphics[scale=0.30]{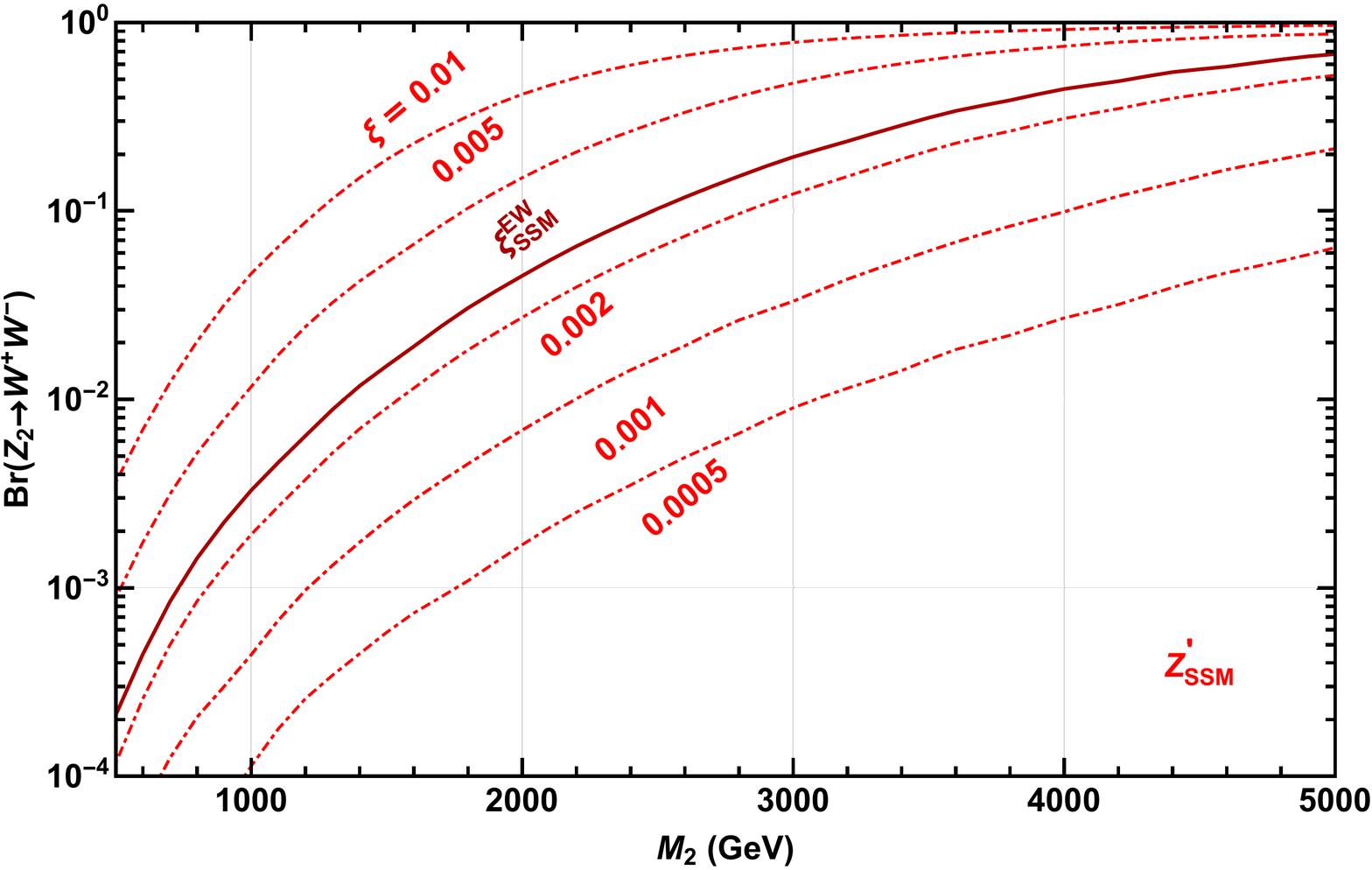}
\end{center}
\vspace*{-5mm}
\caption{ Same as in Fig.~\ref{fig2} but for the ${\rm SSM}$ model
with $\xi_{\rm SSM}^{\rm EW}=2.6\cdot 10^{-3}$.   }
\end{figure}

The total width $\Gamma_{2}$ of the $Z_2$ boson can be written as
follows:
\begin{equation}\label{gamma2}
\Gamma_{2} = \sum_f \Gamma_{2}^{ff} + \Gamma_{2}^{WW} +
\Gamma_{2}^{Z_1H}.
\end{equation}
The presence of the two last decay channels,  which are often
neglected at low and moderate values of $M_2$, is due to $Z$-$Z'$
mixing.  Note, that the widths of these two bosonic modes $W^+W^-$
and $Z_1H$ do not depend on unknown masses of the final states
such as heavy scalars that may enter in some other exotic diboson
channels which we here ignore. The fermion contribution, $\sum_f
\Gamma_{2}^{ff}$, depends on the number $n_g$ of generations of
heavy exotic fermions which can contribute to the $Z_2$ decay without
phase space suppression. This number is model dependent too, and
introduces a phenomenological uncertainty. For the range of $M_2$
values
 below $\sim 3-4$ TeV,  the dependence of
$\Gamma_2$ on the values of $\xi$ (within its allowed range)
induced by $\sum_f \Gamma_{2}^{ff}$, $\Gamma_{2}^{WW}$ and
 $\Gamma_{2}^{Z_1H}$ is unimportant.
Therefore, in this mass range, one can
 approximate the total width as $\Gamma_{2} \approx \sum_f
 \Gamma_{2}^{ff}$, where the sum runs over SM fermions only.
 The ratios of $\Gamma_{2}^{ff}/M_2$ for the benchmark models
 are summarized in Table~\ref{tab1}. One can appreciate the narrowness
 of the $Z_2$ pole from this Table~\ref{tab1}.

However, for large $Z_2$ masses, $M_2>4$ TeV, there is an
enhancement that cancels the suppression due to the tiny $Z$-$Z'$
mixing parameter $\xi$ \cite{Salvioni:2009mt}. While the
``Equivalence theorem'' \cite{Chanowitz:1985hj} might suggest a
value for $\text{Br}(Z_2\to Z_1H)$ comparable to $\text{Br}(Z_2\to
W^+W^-)$ up to electroweak symmetry breaking
effects and phase-space factors,  the $Z_2Z_1H$ coupling is quite
model dependent \cite{Barger:1987xw,Barger:2009xg}. We take an
approach as model-independent as possible, and  for numerical
illustration show our results in two simple scenarios. In the
first scenario (adopted in the bulk of the paper), we treat the model as effectively having a
suppressed  partial width of $Z_2\to Z_1H$ with respect to that of
$Z_2\to W^+W^-$, i.e. $\Gamma_{2}^{Z_1H}\ll\Gamma_{2}^{WW}$, so
that one can ignore the former. In this case, numerical
results with our treatment will serve as an upper bound on the
size of the signal. The second scenario concerns the situation when
both partial widths are comparable, $\Gamma_{2}^{Z_1H}\simeq
\Gamma_{2}^{WW}$ for heavy $M_2$
\cite{Barger:1987xw,Barger:2009xg,Dib:1987ur}. We will start our
analysis from the first scenario and then make comments on the
second one emphasizing the implication of the decay channel
$Z_2\to Z_1H$ for diboson  resonance searches in the process
(\ref{procWW}) at the LHC. In that latter case, one can expect
 that $\Gamma_{2}$ would be larger, with a suppression
in the branching ratio to $W^+W^-$, and the bounds from LHC (and
the ability for observing the $Z$-$Z'$ mixing effect) would be
reduced.

Notice that for all $M_{2}$ values of interest for LHC the width
of the $Z_{2}$ boson is considerably smaller than the experimental
mass resolution $\Delta M$ for which we adopt  the parametrization
in reconstructing the diboson invariant mass of the $W^+W^-$
system, $\Delta M/M\approx 5\% $, as proposed, e.g., in
\cite{Aaboud:2016okv,Sirunyan:2017nrt}.

The expression for the partial width of the $Z_2\to W^+W^-$ decay
channel can be written as \cite{Altarelli:1989ff}:
\begin{equation}
\Gamma_{2}^{WW}=\frac{\alpha_{\rm em}}{48}\cot^2\theta_W\, M_{2}
\left(\frac{M_{2}}{M_W}\right)^4\left(1-4\,\frac{M_W^2}{M_{2}^2}\right)^{3/2}
\left[ 1+20 \left(\frac{M_W}{M_{2}}\right)^2 + 12
\left(\frac{M_W}{M_{2}}\right)^4\right]\cdot\xi^2. \label{GammaWW}
\end{equation}

The dominant term in the second line of Eq.~(\ref{xsection2}), for
$M^2\gg M_W^2$, is proportional to $(M/M_W)^4\sin^2\theta$ and
corresponds to the production of longitudinally polarized $W$'s,
$Z_2\to W^+_LW^-_L$. This strong dependence on the invariant mass
results in a very steep growth of the cross section with energy
and therefore a substantial increase of the cross section
sensitivity to $Z$-$Z'$ mixing at high $M$. In its turn, for a
fixed mixing factor $\xi$ and at large $M_{2}$ where
$\Gamma_{2}^{WW}$ dominates over $\sum_f \Gamma_{2}^{ff}$
\footnote{Here we follow the first scenario, assuming
$\Gamma_{2}^{Z_1H}=0$.} the total width increases
rapidly with the mass $M_{2}$ because of the quintic
dependence on the $Z_2$ mass of the $W^+W^-$ mode as shown in
Eq.~(\ref{GammaWW}). In this case, the
$W^+W^-$ mode becomes dominant and $\text{Br}(Z_2 \to W^+W^-)\to
1$, while the fermionic decay channels ($\Gamma_2^{ff}\propto M_2$) are increasingly
suppressed.

As was mentioned in Sec.~\ref{sec:I}, for models based on the $E_6$ GUT and
left-right symmetry groups, the $Z$-$Z^\prime$ mixing angles (and $\xi$) were excluded
at the level of a few per mil \cite{Erler:2009jh}.  These limits on the
mixing parameter were obtained from an analysis of the $Z^\prime$ extended  models
under consideration against available electroweak (EW) precision data
and are summarized in Table~\ref{tab2}.

All these features are demonstrated in Figs.~\ref{fig2}--\ref{fig4}, where we plot
$\text{Br}(Z_2\to W^+W^-)$ vs $M_2$ for various $Z^\prime$ models and mixing factor
$\xi$  ranging from 0.0005 to 0.01. As reference, we also show the branching ratio
corresponding to $\xi=\xi^\text{EW}$, the bound obtained from the electroweak precision data \cite{Erler:2009jh}.
Values of $\xi$ larger than $\xi^{\rm EW}$ are shown only for
illustrative purposes. It should be stressed that the boost of the branching ratio for high values of $M_2$, illustrated in these figures, plays an important role in the following analysis.

We also note that the branching ratios of the different models are ordered in the following manner
\begin{equation}
\text{Br}(Z_2\to W^+W^-)_\text{SSM}<
\text{Br}(Z_2\to W^+W^-)_\text{LRS}<
\text{Br}(Z_2\to W^+W^-)_\chi<
\text{Br}(Z_2\to W^+W^-)_\eta<
\text{Br}(Z_2\to W^+W^-)_\psi.
\end{equation}
This will be reflected in the bounds obtained.

\begin{figure}[htb]
\refstepcounter{figure} \label{fig5} \addtocounter{figure}{-1}
\begin{center}
\includegraphics[scale=0.40]{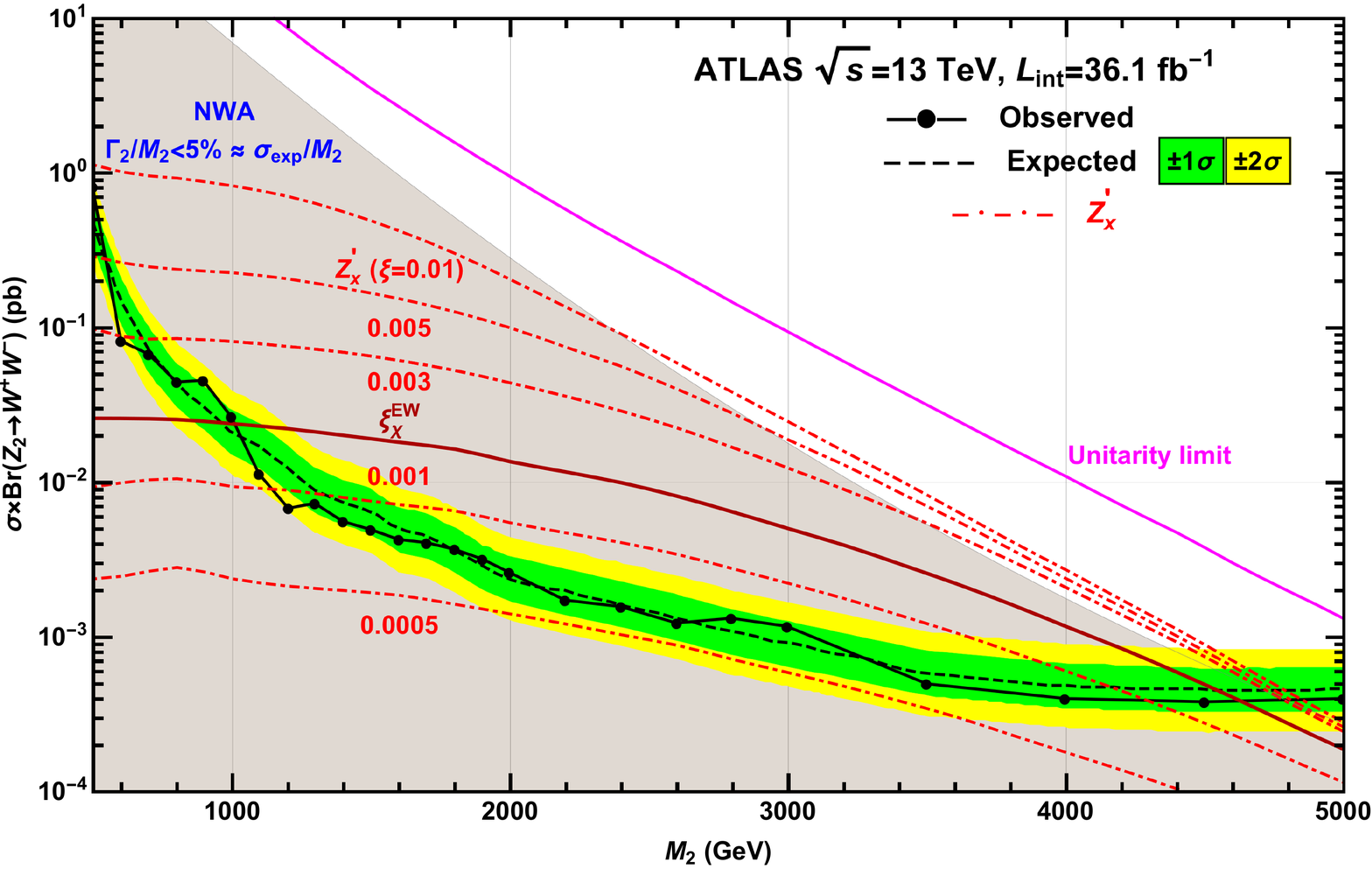}
\includegraphics[scale=0.45]{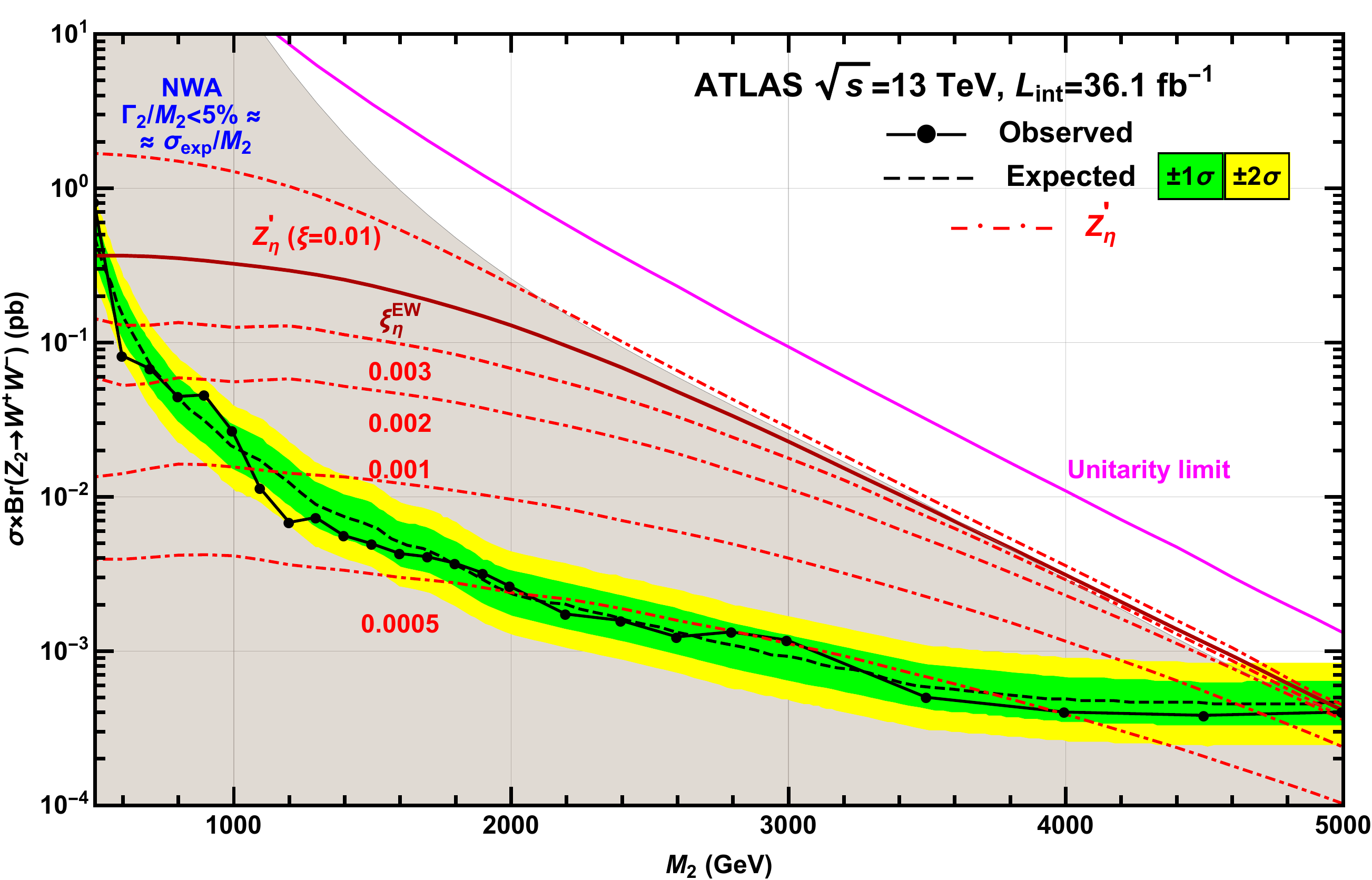}
\end{center}
\vspace*{-4mm}
 \caption{Observed and expected $95\%$ C.L. upper
limits on the production cross section times the branching
fraction for $Z_2\to W^+W^-$ as a function of the $Z_2$ mass, $M_{2}$,
taken from Fig.~7a of Ref.~\cite{Aaboud:2017fgj}, showing ATLAS
data for $36.1~\text{fb}^{-1}$. Theoretical production cross
sections $\sigma\times \text{Br}(Z_2\to W^+W^-)$ for the $\chi$ and
$\eta$ models (upper and lower panels, respectively) are
calculated from PYHTHIA~8.2 with a $K$-factor of 1.9, and given by
dash-dotted curves, for mixing factor $\xi$ ranging from 0.01 and down to
0.0005. Also, the cross section indicated by the solid line
corresponds to the mixing parameter $\xi_{\rm model}^{\rm EW}$.
The area indicated by gray corresponds to the region where the $Z_2$
resonance width is predicted to be less than 5\% of the resonance
mass, in which the narrow-resonance assumption is satisfied.
The lower boundary of the region excluded by the
unitarity violation arguments
is also indicated.
 }
\end{figure}

\begin{figure}[htb]
\refstepcounter{figure} \label{fig6} \addtocounter{figure}{-1}
\begin{center}
\includegraphics[scale=0.45]{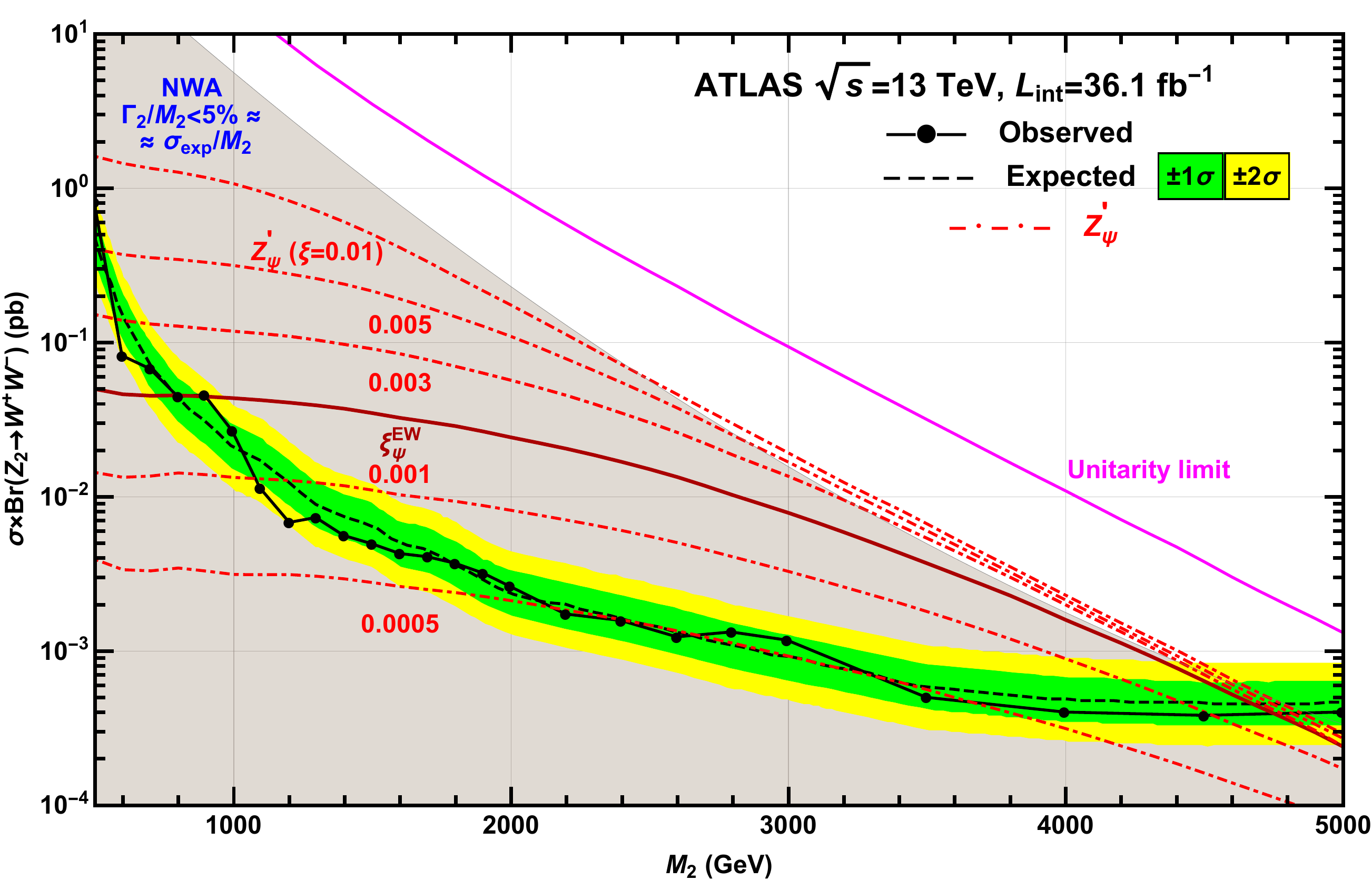}
\includegraphics[scale=0.45]{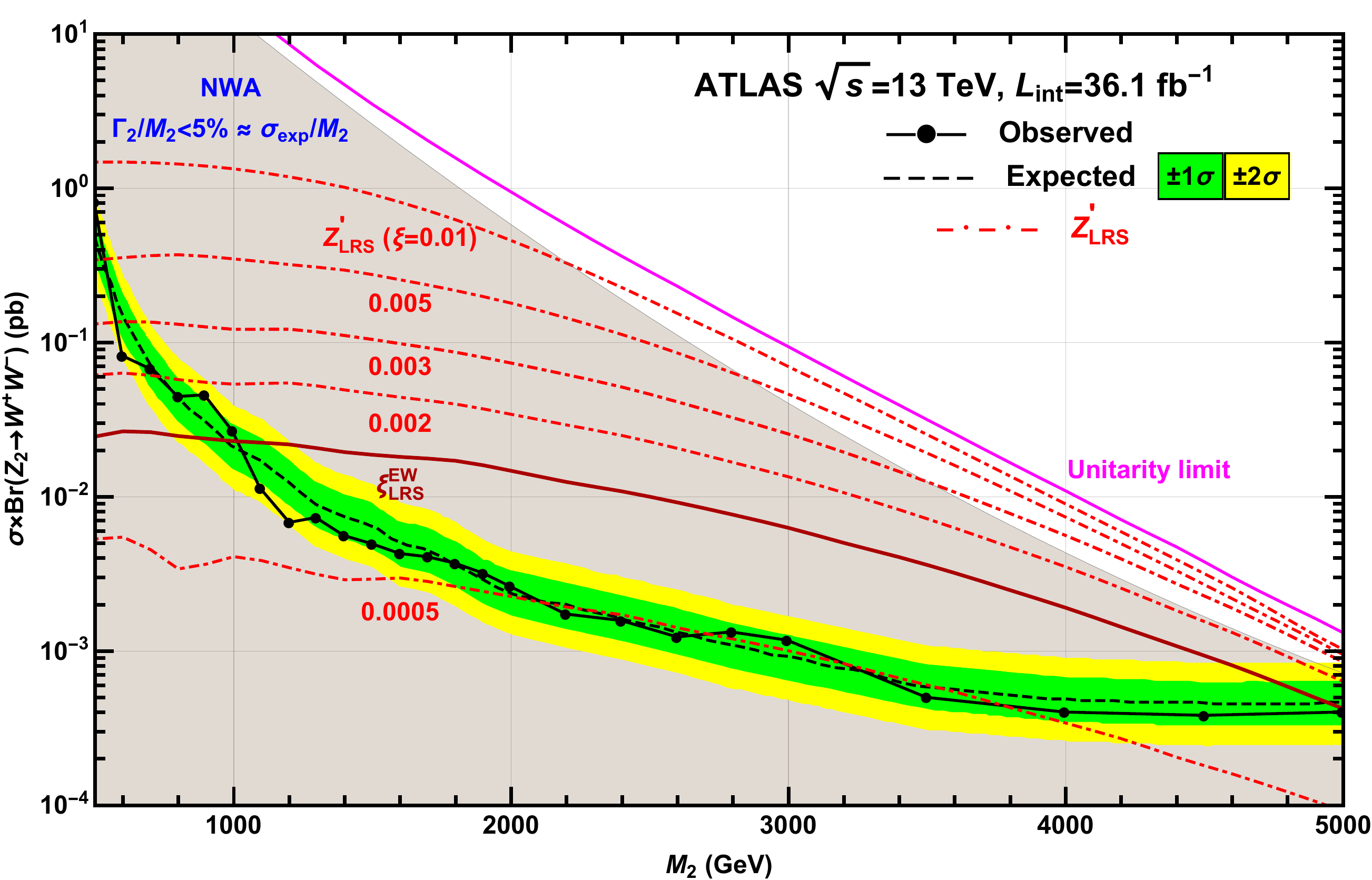}
\end{center}
\vspace*{-4mm}
 \caption{Same as in Fig.~\ref{fig5} but for the the $\psi$ model (upper)
and the ${\rm LRS}$ model (lower panel).
 }
\end{figure}

\begin{figure}[htb]
\refstepcounter{figure} \label{fig7} \addtocounter{figure}{-1}
\begin{center}
\vspace*{-10mm}
\includegraphics[scale=0.45]{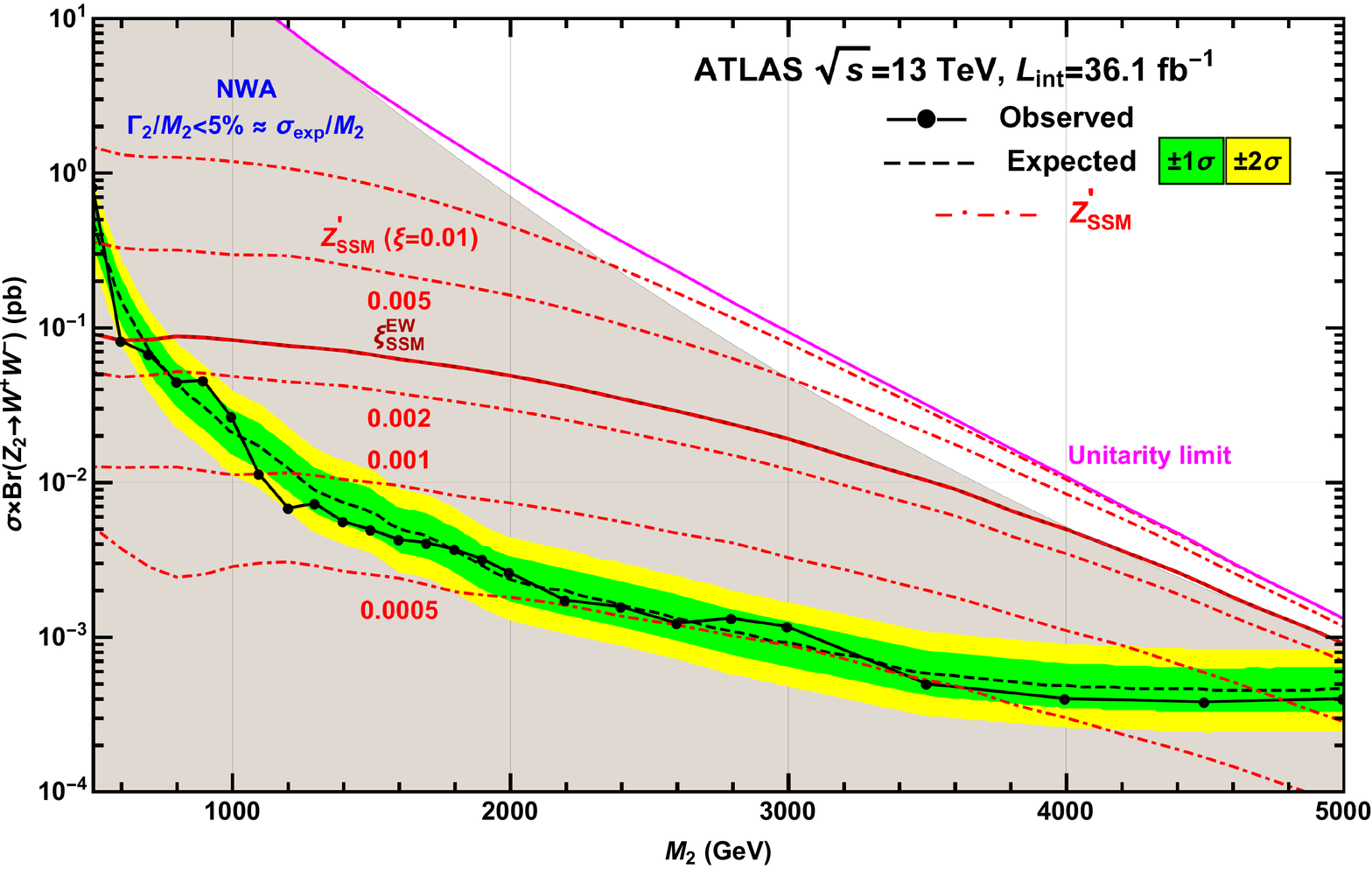}
\end{center}
\vspace*{-4mm}
 \caption{Same as in Fig.~\ref{fig5} but for the ${\rm SSM}$ model.
 }
\end{figure}

\section{Constraints from the diboson process}
\label{sect-constraints}

Here, we present an analysis, employing the most recent
measurements of diboson processes provided by ATLAS
\cite{Aaboud:2017fgj}. We show in
Figs.~\ref{fig5}--\ref{fig7} the observed and expected $95\%$
C.L. upper limits on the production cross section times the
branching fraction, $\sigma\times \text{Br}(Z_2\to
W^+W^-)_{95\%}$, as a function of the $Z_2$ mass, $M_2$. The data
analyzed comprises $pp$ collisions at $\sqrt{s}=13$ TeV, recorded
by the ATLAS (36.1 fb$^{-1}$) detector \cite{Aaboud:2017fgj} at
the LHC. As mentioned above, ATLAS \cite{Aaboud:2017fgj} analyzed
the $W^+W^-$ production in the process (\ref{procWW}) through the
semileptonic final states.

Then, for
$Z_2$ we compute the LHC production cross section multiplied
by the branching ratio into two $W$ bosons, $\sigma \times {\rm
Br}(Z_2\to W^+ W^-)_{\rm theory}$, as a function of the two parameters
($M_2$, $\xi$), and compare it with the limits established by the
ATLAS experiment, $\sigma \times {\rm Br}(Z_2\to W^+ W^-)_{95\%}$.
Our strategy in the present
analysis is to adopt the SM backgrounds that have been carefully
evaluated by the experimental collaborations and simulate only
the $Z_2$ signal.

In these figures, the inner (green) and outer (yellow) bands around the expected
limits represent $\pm 1\sigma$ and $\pm 2\sigma$ uncertainties,
respectively. The theoretical production cross sections
$\sigma\times \text{Br}(Z_2\to W^+W^-)_{\rm theory}$ for $Z_2$
bosons of the benchmark models, are calculated from PYHTHIA 8.2
\cite{Sjostrand:2014zea} adapted for such kind of analysis.
Higher-order QCD corrections to the signal were
estimated using a $K$-factor, for which we adopt a
mass-independent value of 1.9
\cite{Frixione:1993yp,Agarwal:2010sn,Gehrmann:2014fva}. These
theoretical curves for the cross sections, in descending order,
correspond to values of the $Z$-$Z'$ mixing factor $\xi$ from 0.01
to 0.0005. The intersection points of the expected (and measured)
upper limits on the production cross section with these
theoretical cross sections for various $\xi$ give the
corresponding lower bounds on ($M_2$, $\xi$), to be summarized
in Sec.~\ref{sect:overall_constraints}.

The signature space depicted in Figs.~\ref{fig5}--\ref{fig7} is
limited by the assumption that the resonance sought is narrow. The
shaded area represents the region where the theoretical width
$\Gamma_2$ is smaller than the experimental resolution $\Delta M$
($\equiv\sigma_{\rm exp}$) of the searches, and thus where
the narrow-resonance assumption is satisfied. This region
is defined by a predicted resonance
width, relative to its mass, of at most 5\%, corresponding to the best
detector resolution of the searches.

In addition, in Figs.~\ref{fig5}--\ref{fig7} we plot curves
labelled  ``Unitarity limit'' that correspond to the unitarity
bound (see, e.g. \cite{Alves:2009aa} and references therein, where
it was shown that the saturation of unitarity in the elastic
scattering $W^+W^-\to W^+W^-$ leads to the constraint
$g_{Z'WW_\text{max}}=g_{ZWW}\cdot (M_Z/\sqrt{3}M_{Z'})$). This
constraint was adopted in plotting the unitarity bound. It was
obtained under the assumption that the couplings of the $Z'$ to
quarks and to gauge bosons have the same Lorentz structure as
those of the SM, but with rescaled strength.

The signature space displayed in Figs.~\ref{fig5}--\ref{fig7}
bounded by the curve labelled $\xi^\text{EW}$ and the curve
corresponding to the 95\%~C.L. upper limits, $\sigma \times
\text{Br}(Z_2\to W^+W^-)_{95\%}$, is excluded by the ATLAS
experiment. It is interesting to note that for some range of
mixing parameters $\xi$ the $Z_2$ mass may be excluded up to
approximately 5~TeV at 95\%~C.L., which slightly exceeds the
sensitivity of the DY process.

\begin{figure}[!htb]
\refstepcounter{figure} \label{figDY} \addtocounter{figure}{-1}
\begin{center}
\vspace*{-3mm}
\includegraphics[scale=0.40]{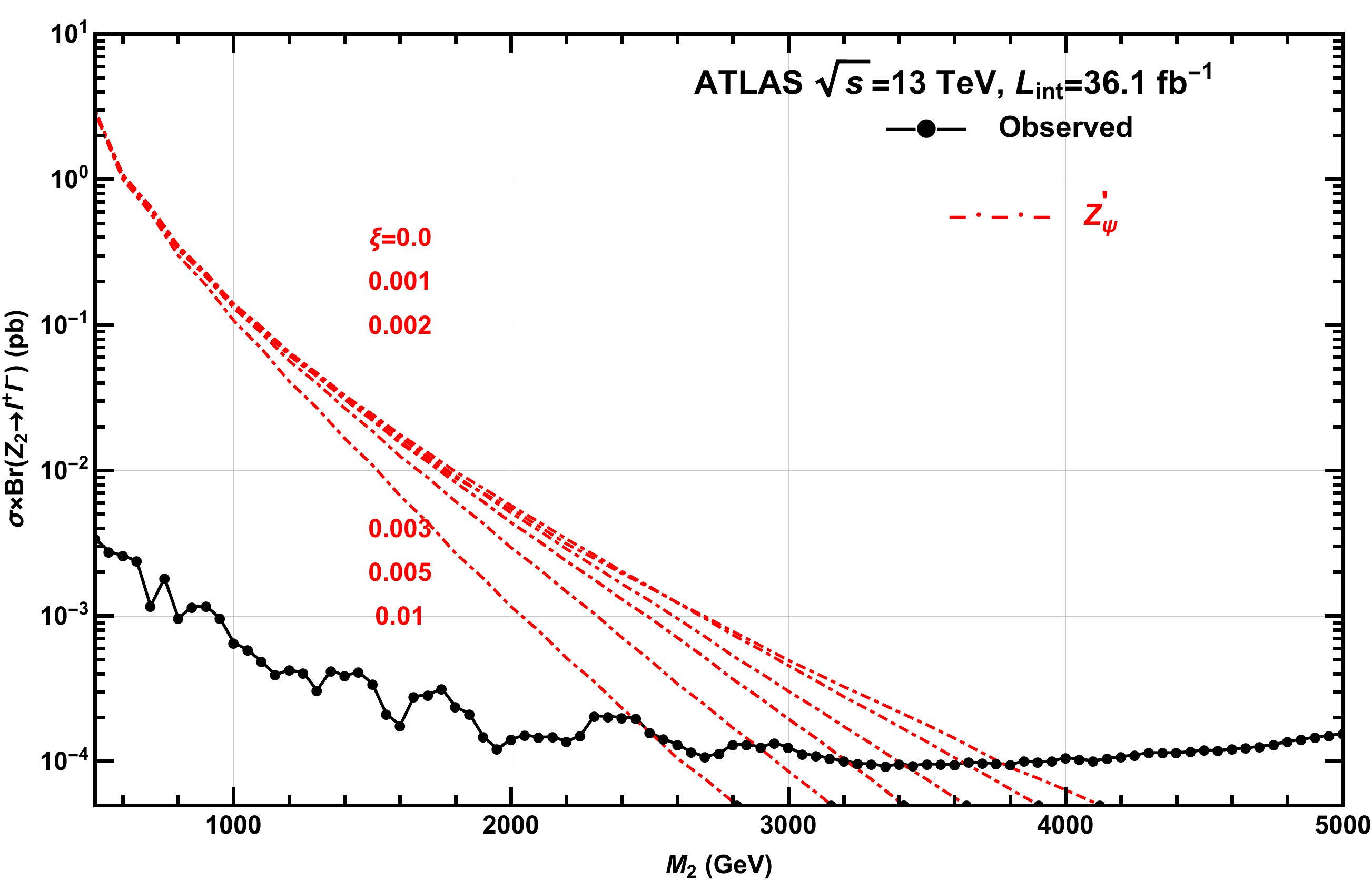}
\end{center}
\vspace*{-4mm} \caption{Solid: observed $95\%$ C.L. upper bound on
the $Z_2$ production cross section times branching ratio to two
leptons, $\sigma\times Br(Z_2\to l^+l^-)_{95\%}$, obtained at the
LHC with integrated luminosity $\Lumint$=36.1\, fb$^{-1}$ by the
ATLAS collaboration \cite{Aaboud:2017buh}. Dash-dotted:
theoretical production cross section $\sigma\times
\text{Br}(Z_2\to l^+l^-)_{\rm theory}$ for the $\psi$ model $Z_2$
boson, calculated from PYHTHIA 8.2 with a $K$-factor of unity.
These curves in descending order correspond to values of mixing
factor $\xi$ from 0 to 0.01.}
\end{figure}

\section{Constraints from the Drell--Yan process}
\label{sect-DY}

The above analysis was for the diboson process (\ref{procWW}),
employing the most recent ATLAS measurements \cite{Aaboud:2017fgj}.
Next, we turn to the Drell--Yan process, this process gives valuable complementary information.
We compute the
$Z_2$ production cross section at the LHC, $\sigma$, multiplied by
the branching ratio into two leptons, $l^+l^-$ ($l=e,\mu$), i.e.,
$\sigma \times {\rm Br}(Z_2\to l^+ l^-)_{\rm theory}$, as a function of $M_2$,
and compare it with the upper limits established by the
experiment \cite{Aaboud:2017buh} for $36.1~\text{fb}^{-1}$.
Results for
$\sigma \times {\rm Br}(Z_2\to l^+ l^-)_{95\%}$ are shown in
Fig.~\ref{figDY}. To account for
next-to-next-to-leading order (NNLO) effects in the QCD strong
coupling constant, the leading
order (LO) cross sections calculated with PYHTHIA 8.2
\cite{Sjostrand:2014zea} are multiplied by a mass-independent
$K$-factor. The value of the $K$-factor is estimated at a dilepton
invariant mass of $3.0-4.5$ TeV and found  to be consistent with
unity \cite{Aaboud:2017buh, Sirunyan:2018exx}.

For illustrative
purposes we show theoretical production cross sections
$\sigma\times \text{Br}(Z_2\to l^+l^-)_{\rm theory}$ for the $Z_2$
boson for only one representative model, $\psi$,  given by the
dash-dotted curves in Fig.~\ref{figDY}. These curves, in descending
order correspond to values of the mixing factor $\xi$ from 0.0 to 0.01.
Qualitatively, the decrease of the theoretical cross section with
increasing values of $\xi$ can be understood as follows: For
increasing $\xi$, the $Z_2\to W^+W^-$ mode will at high mass $M_2$
become more dominant
(as illustrated in Figs.~\ref{fig2}--\ref{fig4}), and $\text{Br}(Z_2\to l^+l^-)$
will decrease correspondingly.
Notice also, that applying a mass dependent $K$-factor
(which for this process is less than 1.04), the $\psi$
model mass limit of the $Z_2$ changes by only $\sim{\cal O}$(50 GeV),
justifying the use of the simpler mass-independent $K$-factor
\cite{Aaboud:2017buh, Sirunyan:2018exx}.

Comparison of $\sigma \times {\rm Br}(Z_2\to l^+ l^-)_{\rm
theory}$ vs $\sigma \times {\rm Br}(Z_2\to l^+ l^-)_{95\%}$
displayed in Fig.~\ref{figDY} allows us to read off an allowed mixing
for a given mass value, higher masses are allowed for smaller mixing, for the reason stated above.
This analysis, illustrated here
for the $\psi$ model can also be
performed for the other benchmark models under
consideration. That comparison can be translated into constraints on the two-dimensional
$M_2$-$\xi$ parameter plane, as will be shown in the next section.

\begin{figure}[htb]
\vspace*{-10mm}
\includegraphics[width=0.42\textwidth]{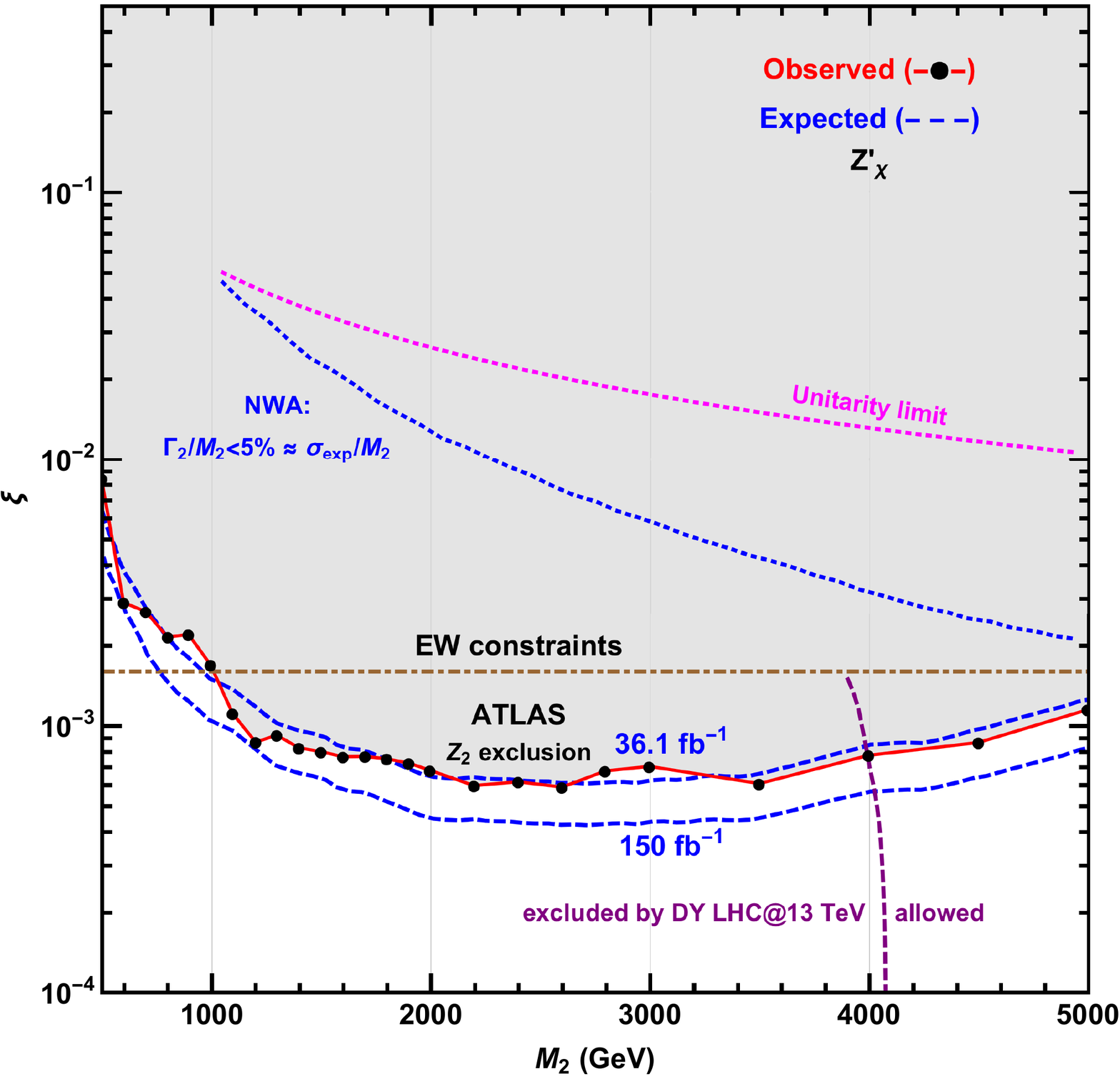}
\includegraphics[width=0.42\textwidth]{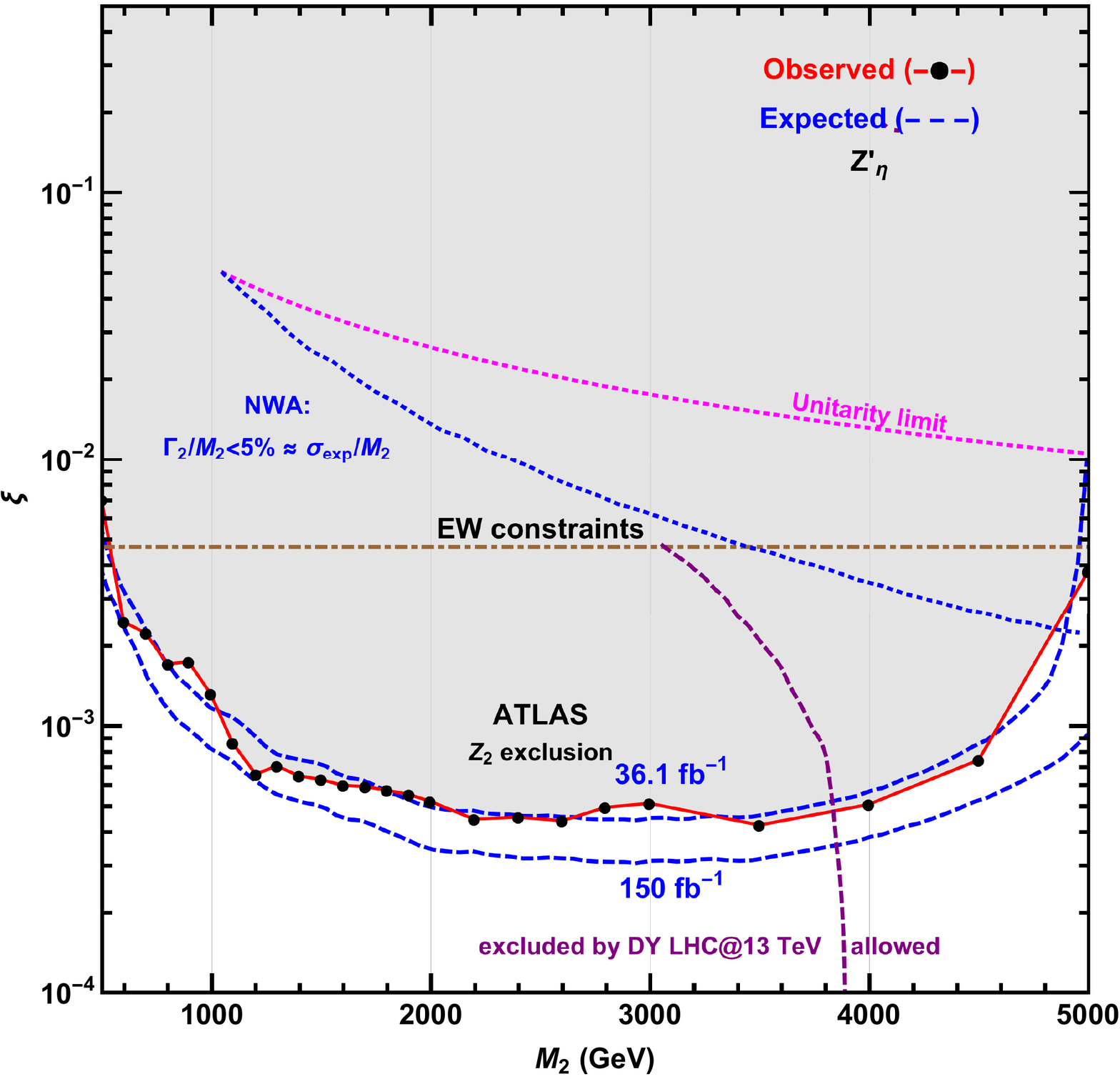}
\vspace*{-10mm}
\caption{\label{fig-limits-a} 95\%C.L. exclusion
regions in the ($M_{2}$, $\xi$) plane of  for the $\chi$ model
(left panel) and the $\eta$ model (right panel) obtained from the
diboson process, given by the boundary of the shaded region. Also
shown is the exclusion from the precision electroweak (EW) data
\cite{Erler:2009jh}. The steep curves labelled ``excluded by
DY LHC@13 TeV'' show the exclusion based on the dilepton channel
\cite{Aaboud:2017buh}. The unitarity limit and the upper bound for
validity of the NWA are shown as dashed curves. Finally, we show
an extrapolation of the expected diboson exclusion that may be
achieved with $150~\text{fb}^{-1}$ of data in Run~II.
 }
\end{figure}

\begin{figure}[htb]
\vspace*{-10mm}
\begin{center}
\includegraphics[scale=0.40]{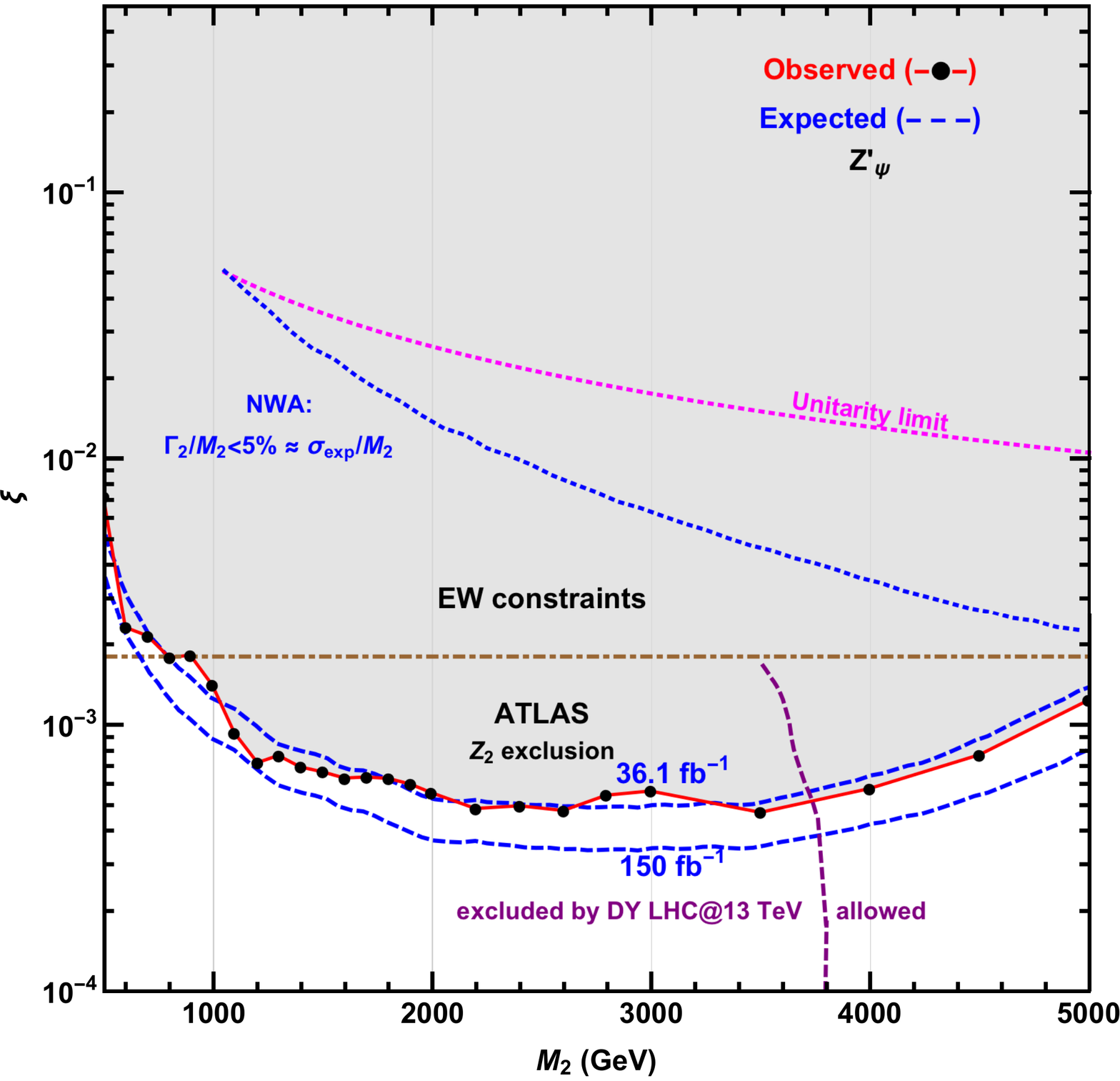}
\includegraphics[scale=0.40]{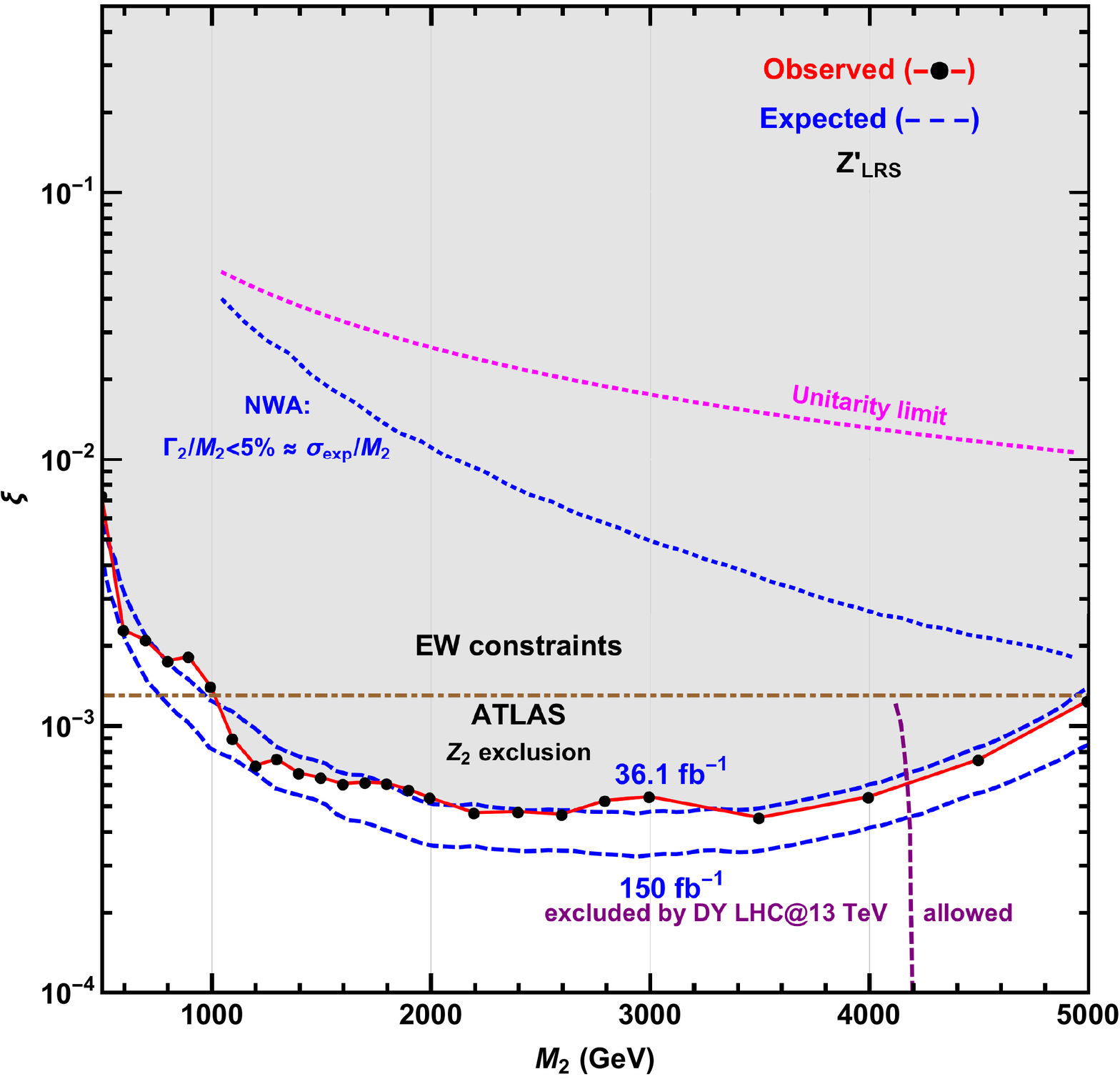}
\end{center}
\vspace*{-15mm}
\caption{Same as in Fig.~\ref{fig-limits-a} but for the $\psi$ model (left
panel) and the ${\rm LRS}$ model (right panel).} \label{fig-limits-b}
\end{figure}

\begin{figure}[htb]
\begin{center}
\vspace*{-15mm}
\includegraphics[scale=0.42]{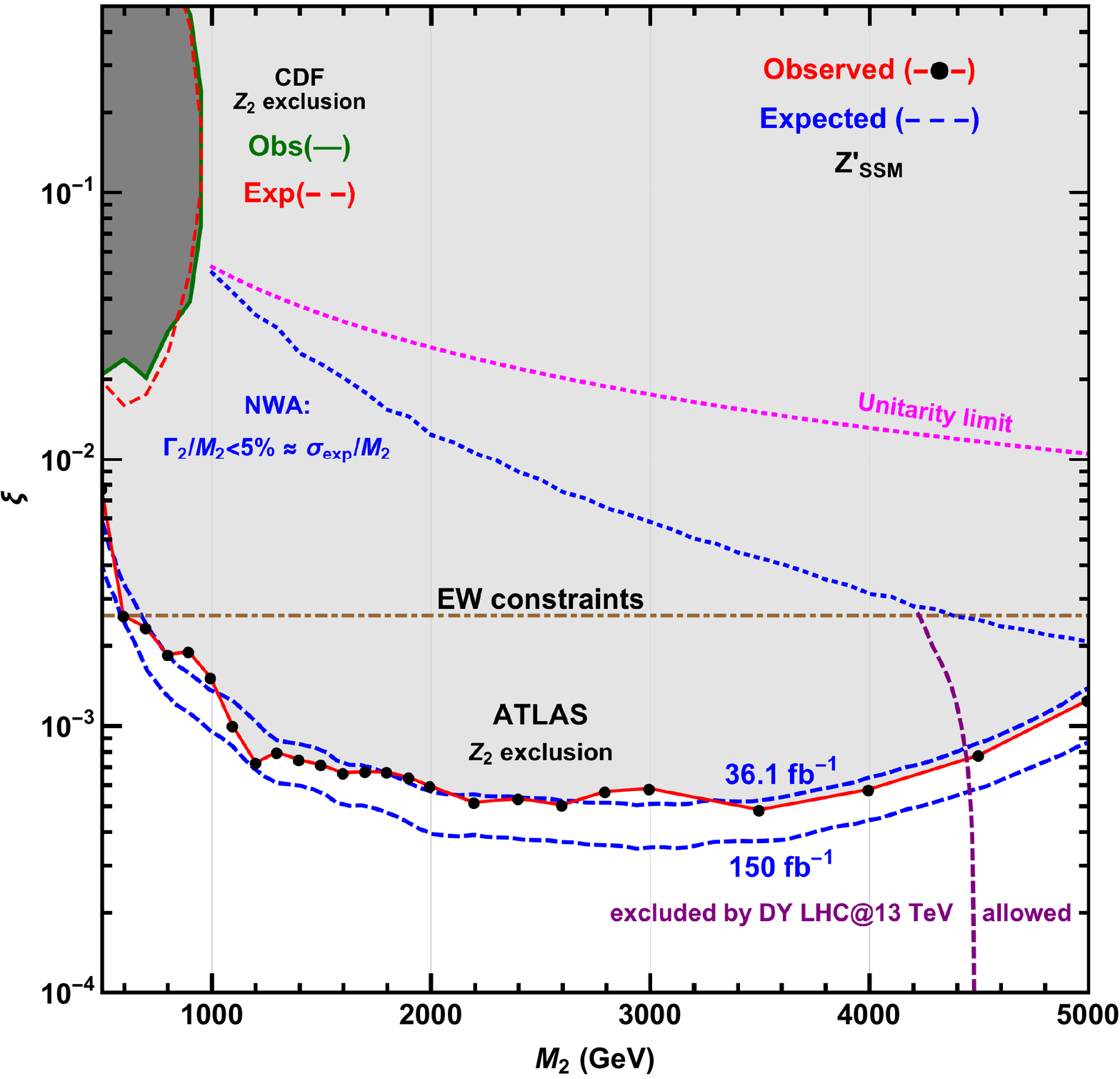}
\end{center}
\vspace*{-15mm}
\caption{Same as in Fig.~\ref{fig-limits-a} but for the ${\rm SSM}$ model.}
\label{fig-limits-c}
\end{figure}

\section{Summarizing constraints on the $Z$-$Z^\prime$ mixing}
\label{sect:overall_constraints}

As described above, both the diboson mode and the Drell--Yan process yield limits on the ($M_2$, $\xi$) parameter space. These are rather complementary, as shown in
Figs.~\ref{fig-limits-a}--\ref{fig-limits-c}, where we collect these and other limits for the considered benchmark
models. The limits arising from the diboson channel are
basically excluding large values of $\xi$, strongest at intermediate masses
$M_2\sim 2-4~\text{TeV}$. The limits arising from the DY channel, on the other hand, basically
exclude masses $M_2\lsim4~\text{TeV}$, with only a weak dependence on $\xi$.
Also, we show the unitarity limits discussed above, as well as the upper bound for the validity of the NWA, both as dashed lines.

Interestingly, these figures show that at
high $Z_2$ masses, the limits on $\xi$ obtained from the ATLAS
diboson resonance production search at 13 TeV are
substantially stronger than those derived from the global analysis
of the precision electroweak data \cite{Erler:2009jh}, which are also displayed. In
Fig.~\ref{fig-limits-c}, which is dedicated to the SSM model, we display limits on the $Z_2$
parameters from the Tevatron
exclusion \cite{Aaltonen:2010ws}, the latter also based on the assumption that no decay
channels into exotic fermions or
superpartners are open.

Furthermore, we have extrapolated the experimental sensitivity
curves for higher expected luminosity downwards by a factor of
$1/\sqrt D$, where $D$ is the ratio of the expected integrated
luminosity of 150 fb$^{-1}$ that will presumably be collected by the end of Run~II,
to the already analyzed integrated luminosity of 36.1~fb$^{-1}$ in
the ATLAS experiment. It is clear that further improvement on the
constraining of this mixing can be achieved from the analysis of
such data. It is easy to see that the exclusion
constraint on $\xi$ at fixed $M_2$ scales as
$\sim\Lumint^{-1/4}$ when statistical errors dominate.
This scaling law $\sim\Lumint^{-1/4}$
for the exclusion bound is an excellent approximation to what is
demonstrated in Figs.~\ref{fig-limits-a}--\ref{fig-limits-c} and in Table~\ref{tab2}.

\begin{table}[htb]
\caption{Constraints on the $Z$-$Z^\prime$ mixing parameter $\xi$
at 95\% C.L.  in different models, processes and experiments.}
\begin{center}
\begin{tabular}{|>{\small}c|>{\small}c|>{\small}c|>{\small}c|>{\small}c|>{\small}c|>{\small}c|>{\small}c| }
\hline collider, process & model & $Z'_{\chi}$ & $Z'_{\psi}$
& $Z'_{\eta}$ & $Z'_{\rm LRS}$ &   $Z'_{\rm SSM}$ & @$ M_2$ (TeV)
 \\
\hline LEP2, $e^+e^- \to W^+W^-$ \cite{Andreev:2012zza} & $\xi
[10^{-2}]$
& 6 & 15 & 50 & 12  & 7 & $\geq 1$\\
\hline Tevatron, $p\bar{p} \to W^+W^- + X$ \cite{Aaltonen:2010ws}
& $\xi [10^{-2}]$
& -- & -- & -- & --  & 2 &  0.4--0.9 \\
\hline electroweak (EW) data \cite{Erler:2009jh} & $\xi^{\rm EW}
[10^{-3}]$
& 1.6 & 1.8 & 4.7 & 1.3  &  2.6 & -- \\
\hline LHC@13~TeV, $W^+W^-$ ATLAS data with 36.1 fb$^{-1}$ (this
work)& $\xi [10^{-3}]$
& 0{.}6& 0{.}5 & 0{.}4 & 0{.}5 & 0{.}4 &  0.5--5.0 \\
\hline LHC@13~TeV, $W^+W^-$ Run~II, (extrap. 150 fb$^{-1}$) (this
work)& $\xi [10^{-3}]$
&  0.4 & 0.3  & 0.3  & 0.3  & 0.3 &  0.5--5.0 \\
\hline \hline ILC@0{.}5~TeV, 0.5 ab$^{-1}$, $e^+e^- \to W^+W^-$
\cite{Andreev:2012cj}& $\xi [10^{-3}]$
& 1{.}5 & 2{.}3 & 1{.}6 & 1.4 &  1{.}2 & $\geq$ 3 \\
\hline ILC@1{.}0~TeV, 1.0 ab$^{-1}$, $e^+e^- \to W^+W^-$
\cite{Andreev:2012cj}& $\xi [10^{-3}]$
& 0{.}4 & 0{.}6 & 0{.}5 & 0.4 & 0{.}3 &  $\geq$ 3 \\
\hline
\end{tabular}
\end{center}
\label{tab2}
\end{table}

In Table~\ref{tab2}, we collect our limits on the $Z_2$ parameters
for the benchmark models. Also shown in Table~\ref{tab2} are the
current limits on the $Z$-$Z'$ mixing parameter $\xi$ from LEP2
and Tevatron, derived from studies of diboson $W^+W^-$ pair
production. The limits on $\xi$ at the Tevatron assume (as does
the present study) that no decay channels into exotic fermions or
superpartners are open to the $Z_2$. Otherwise, the limits would
be moderately weaker. LEP2 constrains virtual and $Z$-$Z'$ boson
mixing effects by the angular distribution of $W$ bosons.
Table~\ref{tab2} shows that the limits on $\xi$ from the EW
precision data are generally competitive with the future collider,
ILC@0.5 TeV, and they are typically stronger than those from the
preceding ``low'' energy colliders such as the Tevatron and LEP2.
 The LHC limits obtained at current c.m.s. $pp$ energy, 13
TeV, and time-integrated luminosity, $\Lumint=36.1$ fb$^{-1}$,
will improve the EW limits by a factor of order 3--10.

\section{Concluding remarks}
\label{sect:conclusions}

The diboson production at LHC@13 TeV allows to place stringent
constraints on the $Z$-$Z'$ mixing angle and $Z_2$ mass, $M_2$. We
derived limits on the mass and the $Z$-$Z'$ mixing angle of the
neutral $Z_2$ bosons by using data from $pp$ collisions at
$\sqrt{s}=13$ TeV and recorded by the ATLAS detector at the CERN
LHC, with integrated luminosity of $\sim$ 36 fb$^{-1}$. By
comparing the experimental limits to the theoretical predictions
for the total cross section of $Z_2$ resonant production and its
subsequent decay into $W^+W^-$ pairs, we show that the derived
constraints on the $Z$-$Z'$ mixing angle for the benchmark models
are of the order of  $\text{a few} \times 10^{-4}$, greatly
improved with respect to those derived from the global analysis of
electroweak data. Further improvement on the
constraining of this mixing can be achieved from the analysis of
data to be collected at higher luminosity expected in Run~II. We
also show that only the future $e^+e^-$ linear collider ILC with
polarized beams and with very high energy and luminosity,
$\sqrt{s}=1$ TeV and $\Lumint=1\, {\rm ab}^{-1}$, may have a
chance to compete with the LHC operating with presently used
energy and luminosity.

Now, let us return to the  issue concerning the second scenario
considered in Sect.~III.  That scenario assumes that the partial
widths are related, $\Gamma_{2}^{Z_1H} = \Gamma_{2}^{WW}$
for heavy $M_2$. Then $\Gamma_{2}$ would be larger by some factor,
with a corresponding suppression
in the branching ratio to $W^+W^-$, and the bounds from the LHC  would
be weaker. However, our calculations show that accounting for the
contribution of the $Z_2$ boson decay channel,  $Z_2\to Z_1H$, to the
total width $\Gamma_2$ does not dramatically change the bounds on
the mixing parameter $\xi$ obtained in the first scenario where
$\Gamma_2^{Z_1H}=0$. Namely, it turns out that the constraints on
$Z$-$Z'$ mixing are relaxed by at most 20-25\% for the higher $Z_2$
masses.

In this paper, for the sake of compactness of the
graphic material, we limited ourselves to an analysis of experimental
data from the ATLAS detector only. Our further analysis shows that the
corresponding CMS data \cite{Sirunyan:2017acf}
yields bounds on the mixing parameter $\xi$ and the $ Z_2 $ boson mass
that agree with the results based on ATLAS data. In addition, our recent
comparative analysis presented in Ref.~\cite{Osland:2017ema}, based on the preliminary
experimental data of the CMS detector at integrated luminosity of
$35.9~\text{fb}^{-1}$ at 13 TeV agree with that performed with ATLAS data,
confirming the equal  sensitivity of the $W$-pair production process
to $Z^\prime$ parameters within the SSM model.
\vspace*{-4mm}

\section*{Acknowledgements}
This research has been partially supported by the Abdus Salam ICTP
(TRIL Programme) and the Belarusian Republican Foundation for
Fundamental Research. The work of PO has been supported by the
Research Council of Norway.


\raggedright

\end{document}